\documentclass{article}

\usepackage{PRIMEarxiv}

\usepackage[utf8]{
inputenc} 
\usepackage[T1]{fontenc}    

\usepackage{hyperref}       
\usepackage{url}            
\usepackage{booktabs}       
\usepackage{amsfonts}       
\usepackage{nicefrac}       
\usepackage{microtype}      
\usepackage{lipsum}
\usepackage{fancyhdr}
\usepackage{graphicx}       
\graphicspath{{media/}}     

\usepackage
{amsmath} 
\usepackage{xcolor}

\usepackage{newfloat}
\usepackage{soul}
\DeclareFloatingEnvironment[name={
Supplementary Figure}]{suppfigure}

\title{
State-switching navigation strategies in \emph{C. elegans} are beneficial for chemotaxis
}

\author{
  Kevin S. Chen $^{\#}$ \\
  Princeton Neuroscience Institute \\
  Princeton University \\
   \And
  Jonathan W. Pillow* \\
  Princeton Neuroscience Institute \\
  Princeton University \\
  \texttt{jpillow@princeton.edu} \\
  \And
  Andrew M. Leifer* \\
  Princeton Neuroscience Institute, Department of Physics \\
  Princeton University \\
  \texttt{leifer@princeton.edu} \\
}

\begin{document}
\maketitle
$^{*}$ Co-senior authors. $^{\#}$ current affiliation: Yale University,  \texttt{kevin.s.chen@yale.edu}
\bigskip 

\begin{abstract}


Animals employ different strategies for relating sensory input to behavioral output to navigate sensory environments, but what strategy to use, when to switch and why remain unclear. 
In \emph{C. elegans}, navigation is composed of ``steering'' and ``turns'', corresponding to small heading changes and large reorientation events, respectively. 
It is unclear whether transitions between these elements are driven solely by sensory input or are influenced by internal states that persist over time. It also remains unknown how worms accomplish seemingly surprising feats of navigation-- for example, worms appear to exit turns correctly oriented toward a goal, despite their presumed lack of spatial awareness during the turn.
Here, we resolve these questions using detailed measurements of sensory-guided navigation and a novel statistical model of state-dependent navigation.
We show that the worm's navigation is well described by a sensory-driven state-switching model with two distinct states, each persisting over many seconds and producing different mixtures of sensorimotor relations. One states is enriched for steering, while the other is enriched for turning. 
This hierarchical, temporal organization of strategies challenges the previous assumption that strategies are static over time and driven solely by immediate sensory input.
Sensory input causally drives transitions between these persistent internal states, and creates the appearance of  ``directed turns.'' Genetic perturbations and a data-constrained reinforcement learning model demonstrate that state-switching enhances gradient-climbing performance.
By combining measurement, perturbation, and modeling, we show that state-switching plays a functionally beneficial role in organizing behavior over time --- a principle likely to generalize across species and contexts.

\bigskip

\textbf{Significance Statement:}

Animals switch between behavioral strategies when navigating sensory environments, but how these decisions are made and whether they benefit navigation remains unclear. We show that \emph{C. elegans} navigate via internal states that persist for  seconds and bias the animal toward either gradual steering or extended bouts of sharp turning. Transitions between these states are sensory-driven. Prior work   hinted at the existence of states and their sensory dependence, but had not  explicitly identified the rules governing the transitions both into and out of these states, or how they benefit performance. We show how sensory-driven transitions between persistent  states benefit gradient climbing and explain an apparent mystery of goal-directed turns. State-switching may be a general feature of natural tasks such as navigation.

\bigskip

\end{abstract}


\keywords{state-space model \and \emph{C. elegans}  \and navigation \and reinforcement learning \and neural circuits}

\section{Introduction}

Animals dynamically alter their behavioral response to sensory stimuli depending on their sensory environment and internal state \cite{Ashwood2022-he, Calhoun2019-ci, Roberts2016-rs, Buchanan_undated-go, liu2022high, priestley2025dorsal}.
Understanding this state-dependence provides insights into behavioral algorithms and neural mechanisms \cite{Roemschied2023-th, Wiltschko2015-uq, Baker2018-ys, Ji2021-il, Berman2016-bh, datta2019computational, flavell2022emergence}.
State-space models have shed light on the state-dependence of fly courtship behavior \cite{Calhoun2019-ci, Roemschied2023-th} and rodent sensory decision-making \cite{Ashwood2022-he, bolkan2022opponent, hulsey2024decision}. But in each of these examples, it is hard to quantitatively relate state-switching to achieving a goal. For example, in fly courtship it is unclear if or how a series of transitioning between ``close,'' ``chase,'' and ``whatever'' states is beneficial for courting a mate.

By contrast, an animal's sensory navigation strategies can often be quantitatively evaluated in terms of the efficiency with which they achieve their navigation goal, e.g., how much each strategy succeeds in collecting reward or approaching the target location.
For instance, flies navigating towards an attractive odor plume exhibit sensory-driven bouts of walking, stopping, or turning, and each motif's contribution to efficient chemotaxis can be clearly quantified \cite{demir2020walking}. But these sensory-guided navigation models have historically omitted internal states, and instead make the simplifying assumption that an animal's response to stimuli is memoryless.

Here we bring these two strands together, by investigating internal states in the context of goal-directed sensory navigation. We do so in the nematode worm \textit{C. elegans}.  The worm's sensory-driven navigation has been extensively characterized \cite{Pierce-Shimomura1999-nt, Iino2009-al, lockery2011computational, Dahlberg2020-ip, Chen2024-od}, including in response to chemicals \cite{Iino2009-al, Pierce-Shimomura1999-nt, Luo2014-pc}, odorants \cite{Iino2009-al, Chalasani2010-cp, Chen2024-od} and temperature \cite{Yamaguchi2018-rh, Atanas2023-bz, ikeda2020context} sensory input, but rarely using internal states \cite{Gordus2015-kk}. Several mysteries about its navigation remain unexplained, however, and they hint at the role of internal states. For example, stateless models of sensory-driven navigation do not explain why \textit{C. elegans} exhibit bouts of turns in quick succession, called pirouettes \cite{Pierce-Shimomura1999-nt, Chen2024-od}, or why they sometimes appear to preferentially exit turns oriented upgradient.

The state-dependent dynamics of locomotion in worms has been extensively characterized in the context of spontaneous behavior without explicitly considering sensory signals \cite{Bartumeus2016-kw, Roberts2016-rs, Stephens2011-de, salvador2014mechanistic, gallagher2013geometry}. 
These analyses are mostly for behavior in open environments without specific task structure \cite{gallagher2013geometry, salvador2014mechanistic, Roberts2016-rs} or focusing on foraging behavior on longer timescales (order of 10 minutes) \cite{Bartumeus2016-kw, calhoun2014maximally}. However, the characterization of behavioral dynamics in a shorter timescale (order of seconds) and how it responds to sensory input to support navigation is still unclear. A rigorous framework to test for state-dependent behavioral dynamics and incorporate sensory-driven navigation is needed in the field.

Here we hypothesize that explicitly considering internal state in \textit{C. elegans} odor-guided navigation may better account for behavioral observations. We test this hypothesis directly by measuring detailed navigation trajectories in worms and inferring strategies through a novel statistical model. We further explore whether switching between states in a sensory-driven manner may provide benefits for efficient goal-directed navigation.

The notion of state is familiar to the \textit{C. elegans} chemotaxis field and is often invoked implicitly or sometimes explicitly, but rarely formalized. For example, the original quantitative description fo \textit{C. eleagns} chemotaxis by Pierce-Shimomura and colleagues describes pirouettes as a ``periods of frequent turning'', which suggests and acts like an internal state. However, with important exceptions \cite{Iino2009-al, Tanimoto2017-pt}, in the quantitative analysis in subsequent works, this notion is not followed up on and turning events are usually considered memoryless and independent \cite{appleby2012model, Dahlberg2020-ip, soh2018computational, Roberts2016-rs, Luo2014-pc, dunn2007circuit, Chen2024-od, Dahlberg2020-ip, ikeda2020context, dunn2007circuit}. Even when included \cite{Iino2009-al, Tanimoto2017-pt}, sensory dependence of transitions are not explicitly described for transition both into and out of such states.
Similarly, on longer timescales when placed off food, there is a notion that the worm transitions from a roaming state to dwelling state \cite{calhoun2014maximally, Bartumeus2016-kw, Flavell2013-pl}. There are even neural correlates in the neuron NSM \cite{ji2021neural, Flavell2013-pl} showing the transition. 
But there has not been explicit quantitative modeling of how sensory inputs drive transition, and there is conflicting evidence as to whether this transition is even sensory-driven \cite{margolis2024stochastic, Roberts2016-rs, Gordus2015-kk}. Here we focus on shorter timescales of sensory-driven navigation and seek to explicitly define sensory-driven states.

We formulate a novel statistical model that explicitly includes state-dependent navigation strategies. We find that the inclusion of sensory-driven states better explains the experimental measurements of worm navigation and that state-switching is beneficial for navigating upgradient. The model recapitulates the original observations surrounding pirouettes and also explains why worms appear to exit turns in the preferred direction. Importantly, this explains directed turns in a way that respects the worm's presumed lack of awareness of its own heading during the turn. By perturbing and ablating individual neurons, we show evidence that suggests that the neural mechanisms that control behavioral kinematics may be distinct from neural mechanisms that control state-dependent navigation. Furthermore, we show that in a more loosely constrained model, sensory-driven state-switching solutions emerge naturally when optimizing for chemotaxis performance. These approaches corroborate the description of state-switching strategy in worms and show that it is functionally beneficial for navigation.

\section{Results}

\subsection{Worms switch between state-dependent navigation strategies in a sensory-driven manner}


\begin{figure}[!t]
\centering
\includegraphics[width=.9\textwidth]{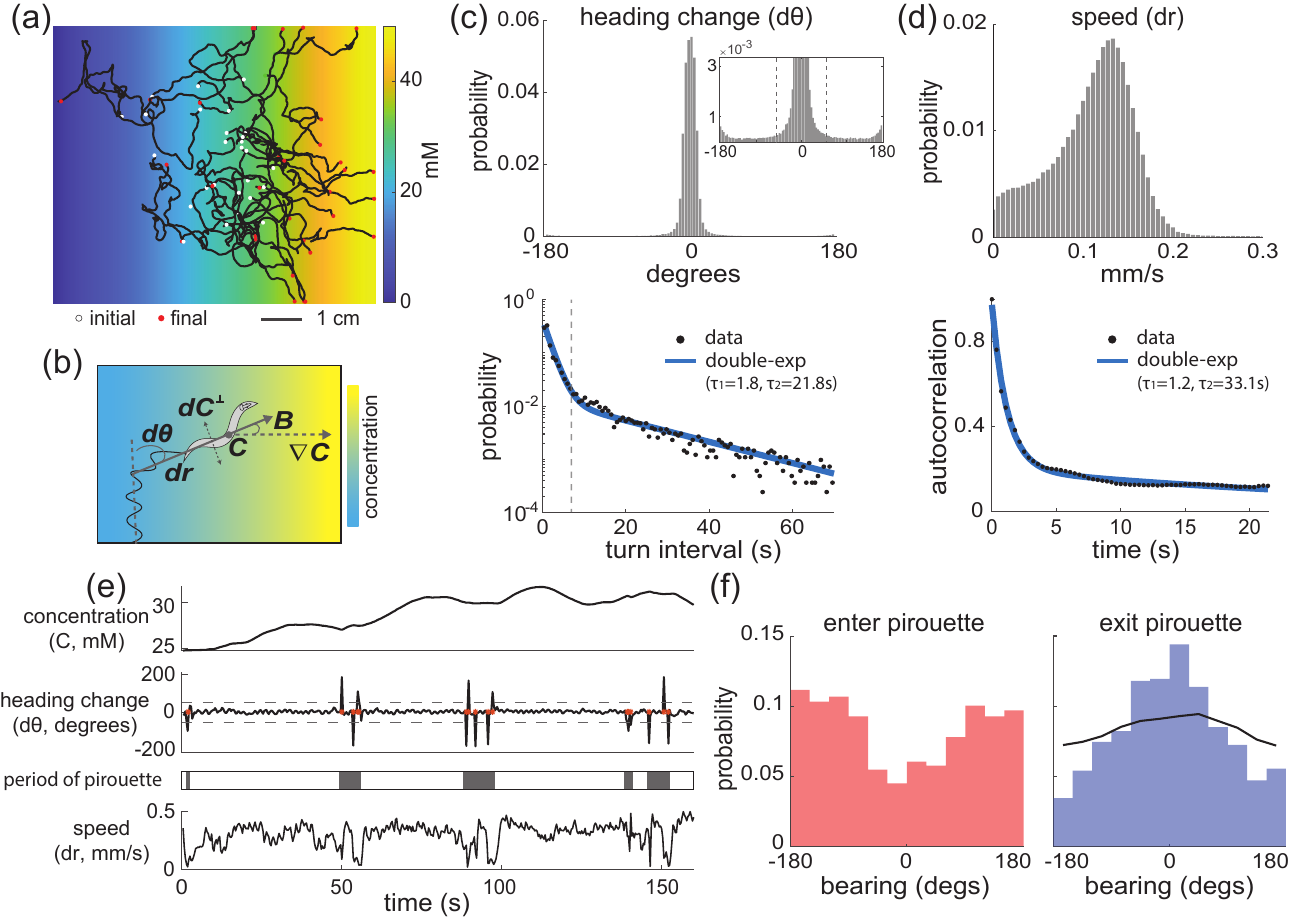}
\caption{\textbf{Behavioral dynamics suggest that the worm may be alternating between persistent states.} \textbf{(a)} Example worm chemotaxis trajectories measured in a linear salt gradient. \textbf{(b)} Observables are extracted from the chemotaxis experiments. The speed $r$, heading change $d\theta$, concentration $C$, perpendicular concentration difference $dC^{\perp}$, and bearing $B$ towards the local gradient $\nabla C$ are shown.
\textbf{(c)} The distribution of heading change $d\theta$ (top) and with inset showing small probability region. The threshold for a turn ($\pm 50$ degrees) is indicated with dash lines. The distribution of turn intervals and the double exponential fit in blue are shown (bottom). The crossing point of the two fitted exponential ($t_c=6.8$s) is shown as a vertical dash line. This value is used to define the period of pirouette in panel (e) in accordance with \cite{Pierce-Shimomura1999-nt}. 
The single-exponential fit has $R^2=-98.3$ and double-exponential fit has $R^2=0.99$.
\textbf{(d)} The distribution of speed $dr
$ (top) and its autocorrelation 
(bottom). The data points are fitted with a double-exponential curve shown in blue.
The single-exponential fit has $R^2=0.72$ and double-exponential fit has $R^2=0.99$. 
\textbf{(e)} Example navigation trajectory measured from experiment, with turns crossing the threshold (dash lines) shown in red dots. The pirouette state indicated in gray is identified following \cite{Pierce-Shimomura1999-nt}.
\textbf{(f)} The bearing before the worm enters a pirouette (left) and after it exits it (right). The calculation of the exiting bearing with shuffled turning angles shown in black line. Quantification in c, d, and f are for $\sim20$ animal-hours of navigation trajectories.
}
  \label{fig:fig0}
\end{figure}

In order to quantify the worms' navigational dynamics, we measured behavioral time series and sensory signal during chemotaxis assays. Specifically, we tracked worms' central positions as they navigated in a linear salt gradient or butanone odor gradient (Fig~\ref{fig:fig0}a) and measured their heading change ($d\theta$) and speed ($dr$) along the navigation trajectories (Fig~\ref{fig:fig0}b). 
In total, we recorded navigation from 7–11 plates per condition, resulting in 127–226 animal-hours across both new and previously published measurements \cite{Chen2024-od}. 
We found several signatures in the statistics of their movements that suggest that worms may be alternating between two locomotory states, which persist in time on the timescale of seconds. 
We inspected the distribution of turning events, where a turn is defined as instantaneous events when the heading change first exceeds a threshold, and observed that the inter-turn interval distribution had two distinct timescales and was well fit by a double but not single exponential (Fig~\ref{fig:fig0}c). This is suggestive of a system with at least two states. 
By contrast, if the animal had only a single state with respect to turning events, we would expect the single exponential to fit well. 
Similarly, the autocorrelation of the animal's speed exhibited two timescales that was well-fit by a double but not single exponential (Fig~\ref{fig:fig0}d). These timescales match what Pierce-Shimomura and Lockery described as ``periods of frequent turning''-- what they called pirouettes \cite{Pierce-Shimomura1999-nt} (Fig~\ref{fig:fig0}e). It has been observed, and we confirmed, that the ``periods of frequent turning'' or pirouettes appear to be related to goal-directed navigation (Fig~\ref{fig:fig0}f). While sensory-driven turning appears in many models of chemotaxis, including biased random walks for bacterial chemotaxis \cite{berg1975chemotaxis}, grouping of turns has been less well explored. We found that animals are more likely to be heading down-gradient from an attractive odorant when they enter a pirouette, and more likely to be heading upgradient when they exit the pirouette (Fig~\ref{fig:fig0}f). In other words, the grouping of turns into pirouette seems to be important for reorienting the animal up gradient.

While it is clear that the animal is able to reorient towards the goal direction, it remains a mystery how it is able to exit a pirouette in the preferred direction-- sometimes called a ``directed turn.'' Directed turns are of interest in the field because the worm is not usually thought capable of maintaining moment-by-moment knowledge of its heading with respect to the gradient during a turn.  We seek to understand how the presence of persistent behavioral states facilitates goal-directed navigation and how it explains directed turns and other features of navigation.

\begin{figure}[!t]
\centering
\includegraphics[width=.84\textwidth]{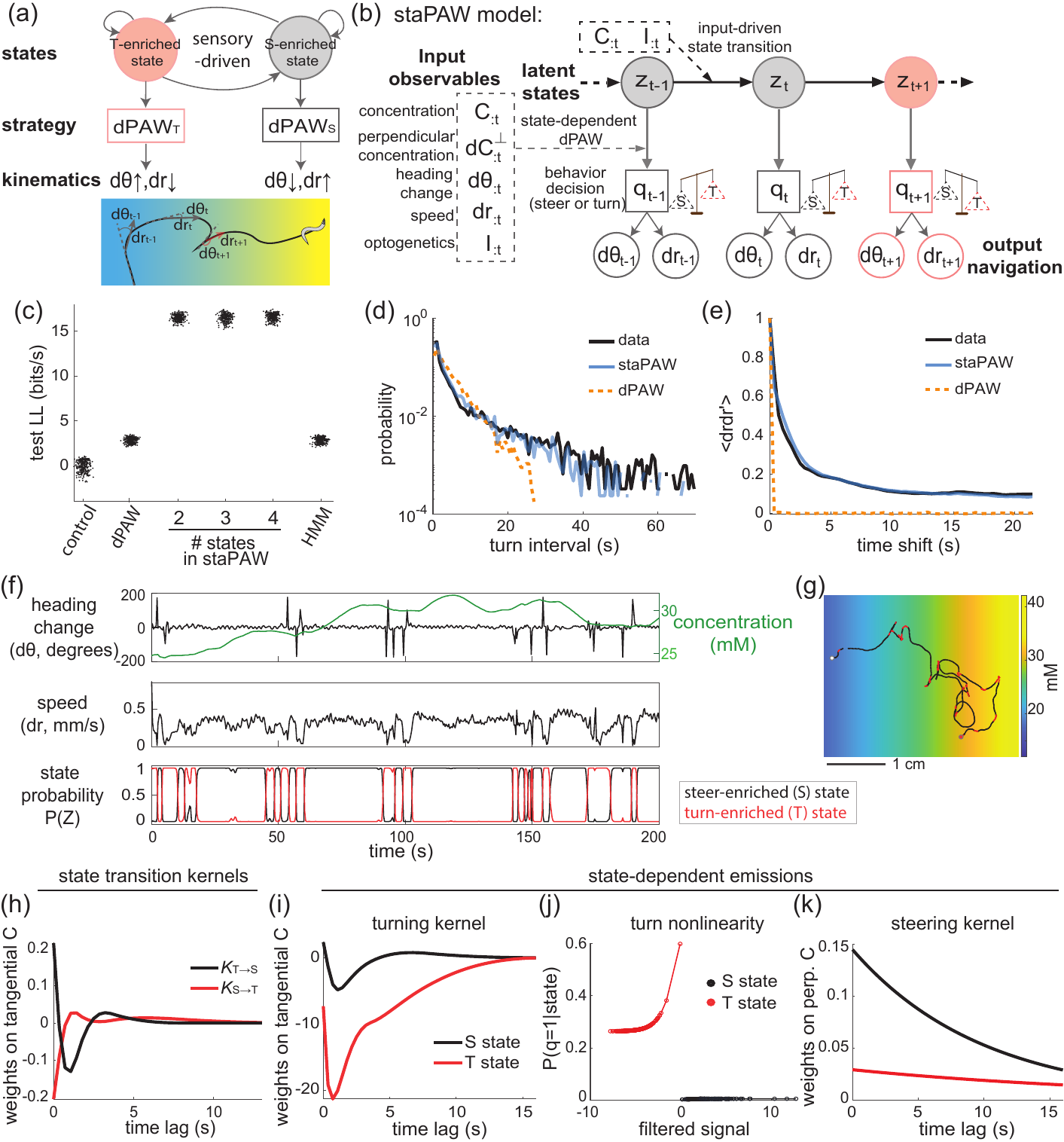}
\caption{\textbf{Two states better characterizes navigation behavior than one state.} 
\textbf{(a)} Schematic of a state-dependent navigation strategy. This is an example with two states, one is the steer-enriched (S) state and another one is the turn-enriched (T) state, each having a dPAW emission for behavioral output. A cartoon time series of worm navigation trajectory color-coded by the behavioral state is shown below. 
\textbf{(b)} Graphical model of staPAW unrolled in time with detailed input-output relation. The input observables are extracted from experimental measurements. The staPAW model assumes a latent discrete state $z$ governing the mapping from input to state-dependent output navigation. 
\textbf{(c)} Test log-likelihood across different statistical models. The difference compared to the mean of control model is plotted on the y-axis. The cross-validation process is 3-fold, each with 5 inference instantiations, and tested with 20 sampled trajectories (with 60 minute total animal time length).
\textbf{(d)} Turn interval distribution from data and prediction from the staPAW or dPAW models. 
\textbf{(e)} Speed autocorrelation from in data and prediction from the staPAW or dPAW models.
\textbf{(f)} Example time series of a two-state staPAW fit to data. The posterior probability of being in SE and TE states are shown below. 
\textbf{(g)} Example navigation trajectory plotted in space. Black and red states correspond to the same hidden state in (f). The white circled dot and red circled dot indicate the initial and final location. 
\textbf{(h)} State transition kernels between two latent states.
\textbf{(i-k)} State-dependent parameters for emission. (i) The turning kernel weights the tangential concentration conditioned on two states. 
(j) The nonlinear function that maps filtered
signal to turn probability (including history angle change and concentration), conditioned on two states. 
(k) The steering kernel weights the perpendicular concentration conditioned on two states.
}
  \label{fig:fig1}
\end{figure}

To investigate the role of states, we developed a novel hierarchical state-switching model of goal-directed navigation (Fig~\ref{fig:fig1}a,b). 
This model resembles a generalized linear model - hidden Markov model (GLM-HMM), where sets of GLMs relate sensory input to behavioral output and the HMM introduces state-dependency and transitions. 
The top level of the hierarchy is the state of the animal that can persist for many seconds. It is composed of latent discrete states, $z$, that follow a Markov chain with transition probability driven by sensory input:
\begin{equation}\label{eq:1}
    P(z_{t+1}=j  | z_t=i , C_{1:t-1}) = \frac{\exp ( K_{ij} \cdot \mathbf{C}_{1:t-1} + b_i ) }{Z} 
\end{equation}
where $K_{ij}$ is the state transition kernel acting on concentration time series $\mathbf{C}_t$ that drive transitions from state $i$ to $j$, $b_i$ is the baseline probability of self-transition, and $Z$ is the normalization constant.

The second layer of the hierarchy captures what has traditionally been called the navigation strategy; here it is a weighted mixture of two sensorimotor strategies \cite{Chen2024-od}: one that invokes individual turns as reorientation events to form a biased random walk \cite{berg1975chemotaxis} and one based on weathervaning in which the animal gradually alters its heading \cite{Iino2009-al, Hendricks2012-bw}.  The choice between these two strategies is a sensory-dependent Bernoulli process that decides to produce an individual turn ($q=1$) or continue to weathervane ($q=0$). Within a state $z$, the decision follows: 
\begin{equation}\label{eq:2}
    P(q_t=1  | z_t ) = m^z+ \frac{M^z-m^z}{1+\exp( g( \mathbf{C}_{1:t-1},\mathbf{d\theta}_{1:t-1} | z_t))} 
\end{equation}
where the state-dependent parameters are denoted with superscript $z$, with the maximum $M$, minimum turn probability $m$, and contributions from past concentration $\mathbf{C}_t$ and behavior $\mathbf{d\theta}_t$. The detailed calculation of function $g$ is described in Methods.

The lowest layer of the hierarchy is the kinematics produced by the animal, such as the speed of its movement ($dr$) or the change of its heading ($d\theta$). Conditioned on being in one state $z$, the kinematics follow:
\begin{equation} \label{eq:3}
    P(d\theta_{t}, dr_t | C_{1:t-1}, dC_{1:t-1}^{\perp}, d\theta_{1:t-1} ; z_t) = 
    P(q_t=1 | z_t)P_{\mathrm{tr}}(d\theta,dr | z_t)
    + P(q_t=0 | z_t)P_{\mathrm{wv}}(d\theta,dr \mid dC^{\perp}_{1:t-1}, z_t)
\end{equation}
which forms a model with mixture probability of turns $P_{\mathrm{tr}}$ and weathervaning $P_{\mathrm{wv}}$ behavioral strategies. 
These kinematics depend on the experienced tangential concentration $C$, perpendicular concentration difference $dC^{\perp}$, and history of behavior $d\theta$. The detailed parameterization of distributions $P_{\mathrm{tr}}$, $P_{\mathrm{wv}}$, and dependence on sensory input are given in Methods.

Without the top layer the model reverts to a stateless model that we studied previously called dPAW, which assumes navigation is memoryless beyond the timescale of a few seconds \cite{Chen2024-od}. We therefore call the complete hierarchical framework a state-dependent Pirouettes and Weathervaning model, or staPAW.

Including states improves the model's agreement with experimental measurements of worm navigation. Specifically, the staPAW model with state-dependency improves the log-likelihood of fit to a held-out test dataset (Fig~\ref{fig:fig1}c), and recapitulates the measured distribution of turning intervals (Fig~\ref{fig:fig1}d) and speed autocorrelations (Fig~\ref{fig:fig1}e) compared to the state-less dPAW model. Furthermore, we used the fitted model to predict time-varying behavioral output (SFig~\ref{fig:SI1}). In general, it is challenging to predict the precise timing of behavior events, such as turns, that are assumed to be stochastically generated.  Nonetheless, the state-based model staPAW outperformed the stateless dPAW in predicting the time series of turning events and speed. 
The staPAW model fitted to measurements reveal the spatiotemporal structure of states during navigation (SFig~\ref{fig:SI_states}). Additionally, once fitted, the generates navigation trajectories that match observations (SFig~\ref{fig:SI2}). 
Together, this indicates that staPAW captures the time-varying and state-switching sensory navigation strategies in worms better than a stateless model.

We wondered what conferred the staPAW model its ability to better capture the properties of measured navigation. We first investigated how performance varied with the number of states. We tested variants with differing numbers of states, ranging from the stateless dPAW model to a staPAW variant with four states. We computed their performance in a cross-validated manner using held-out data. Including state-dependency dramatically outperforms the stateless model,  (Fig~\ref{fig:fig1}c) but adding additional states beyond the two-state model did not significantly increase performance.

The full two-state model assumes that state-switching is driven by sensory input. We tested whether sensory input-driven state-switching confers a performance benefit compared to merely having two states that transition irrespective of sensory input. 
We studied a variant of staPAW that can only have Markov transitions between states and is therefore independent of sensory input. This is equivalent to a Hidden-Markov model (HMM). The HMM model performed dramatically worse than a sensory-driven state-switching model and similarly to stateless dPAW. This suggests that it is specifically sensory-driven state-switching, and not merely the presence of states that confer the model's performance. 

All models, even the HMM, take into account sensory input at the strategy level to inform the generation of kinematics. As expected, all models outperform a stateless control null model that has no access to any sensory input.



\begin{figure}[!t]
\centering
\includegraphics[
width=.6\textwidth
]{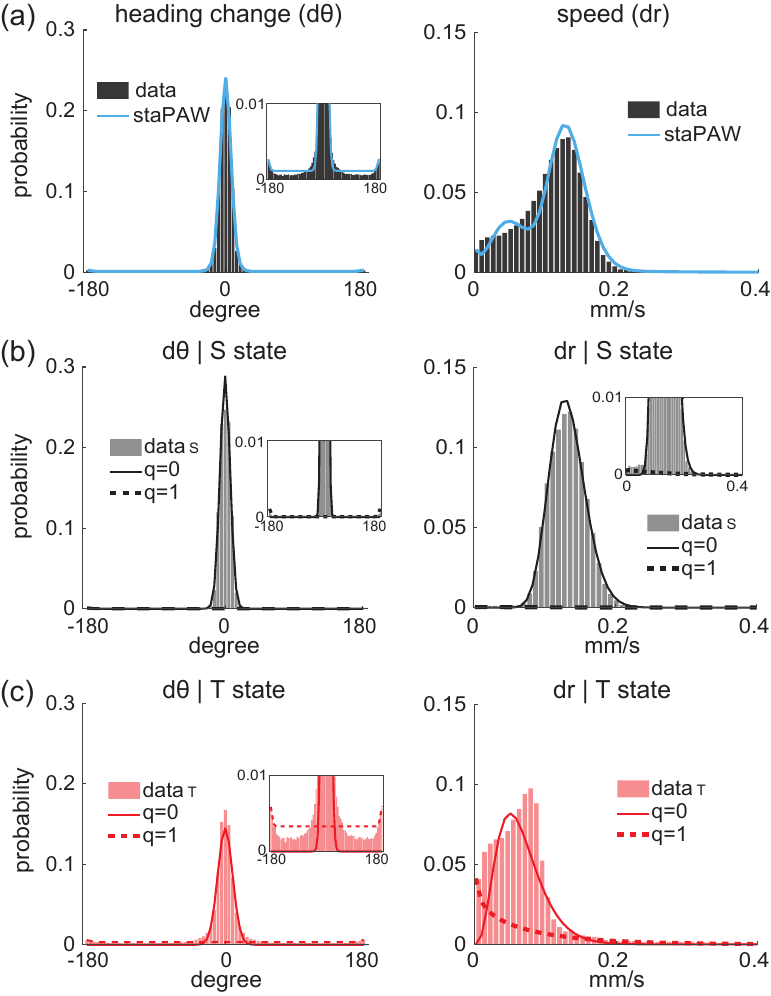}
\caption
{\textbf{One state is enriched for steering and another state is enriched for turning.} 
\textbf{(a)} The distribution of heading change ($d\theta$) on the left and for speed ($dr$) on the right. Measured data is compared to prediction from staPAW.
\textbf{(b)} Same representation as in (a) but conditioned on S-state.
\textbf{(c)} Same representation as in (a) but conditioned on T-state.
Within each state, the predicted distribution corresponding to the turn decisions $q=0$ and $q=1$ are shown.
}
  \label{fig:fig2}
\end{figure}

\subsection{Two behavioral states correspond to a turn-enriched and a steer-enriched state}

What do the two states represent in the two-state staPAW?
The fitted staPAW model reveals that worms switch between two distinct behavioral states during navigation (Fig~\ref{fig:fig1}f,g) --- what we call a steer-enriched state (S) and a turn-enriched state (T).  In the course of navigation, the animal switches between the S and T-states (Fig~\ref{fig:fig1}f). During the S-state the animal has few sharp turns and moves at relatively higher speed , while in the turn-enriched T-state the animal has more frequent sharp turns at lower speed (Fig~\ref{fig:fig2}). The mean time in T-state is 4.6 seconds (with on average 1.9 reversals per transition) and 8.1 seconds in the S-state. These results are superficially similar to the previous report of pirouette-and-weathervaning behavior in worm chemotaxis \cite{Pierce-Shimomura1999-nt}, but here we show quantitatively that state-switching better matches observed behavior. Moreover, for the first time we present a formal mathematical framework that describes how sensory input drives switching between states.

The fitted parameters in staPAW reflect how the worm's transition between states depends on concentration input during navigation. At the state transition level, we found that the state transition kernel for two directions (state T$\rightarrow$S and S$\rightarrow$T) are close to mirror images (flipped-sign) of each other (Fig~\ref{fig:fig1}h).
$K_{S\rightarrow T}$ resembles a negative derivative kernel of the chemical concentration and therefore is consistent with the biased-random walk description -- when life is getting worse (attractive chemical is decreasing) it is time to reorient \cite{Pierce-Shimomura1999-nt, berg1975chemotaxis}.  Unlike the traditional biased random walk, in this state-based formulation the reorientation is not just a single turning event but rather a transition into a turn-enriched state composed of multiple turns. 

Interestingly, $K_{T \rightarrow S}$ resembles a positive derivative kernel, which indicates that the worm is more likely to exit the T-state and enter the S-state when the chemical concentration increases. This adds an additional logic element to the traditional biased random walk framework. The organism not only transitions into a turning enriched state when life is getting worse, when life is getting better it also biases its transitions out of that state. We will discuss in the following section that this is important for explaining the non-uniform distribution of bearing angle when exiting the turning state, and the associated mystery of the worm's apparent directed turn (Fig~\ref{fig:fig0}f).

The T $\leftrightarrow$ S kernels discussed above describe the sensory-driven transitions between states. Within a state, there are also parameters such as the turning kernel
that describe how the tangential concentration influences the production of individual turns, and these also differ between the T and S-states.  In the S-state, these kernels resemble that of a derivative, whereas in the T-state the kernel includes a strong negative integration component (Fig~\ref{fig:fig1}i). The negative shape of these kernels indicates that the turn rate increases when the tangential concentration decreases.

The tangential concentration convolved with its turning kernel yields a filtered signal \cite{Chen2024-od}.  A nonlinear function describes how the filtered signal evokes a turn. The nonlinearities reveal that in the T-state the turn rate is elevated and depends tightly on the filtered signal, while in the S-state the turning rate is consistently low and independent of the filtered signal (Fig~\ref{fig:fig1}j).  In other words, the worm modulates turning based on concentration in the  T-state, but suppresses turning in the S-state.

Conversely, the worm suppresses steering during T-state but steers its heading according to the perpendicular concentration during the S-state. This can be seen in the elevated increased steering kernel in the S compared to T-state (Fig~\ref{fig:fig1}k). These state-dependent kernels reveal how worms navigate dynamically in the sensory environment depending on the time-varying S or T-states.

At the level of kinematics (top row), the animal's behavior varies significantly across the two states (Fig~\ref{fig:fig2}).  In the S-state (middle row) the worm rarely produces turns (where $q=1$ denotes turn), and instead only alters its heading $d\theta$ by small changes ($q=0$). In contrast, in the T-state (bottom row) the probability of making a turn ($q=1$) is much larger, resulting in a much broad distribution of heading changes. Similarly we found increased runs (higher $dr$) in the S-state and decreased runs (lower $dr$) in the T-state.  By decomposing the animal's navigation into two states S and T, each containing both turns ($q=1$) and non-turns ($q=0$),  the model shows excellent agreement to measurements of navigation.

The timescale between instances of  S-states (SFig~\ref{fig:SI_states}a), which emerges naturally from fitting the model to measurements, is broadly consistent with previously described heuristics in the literature used to identify pirouettes \cite{Pierce-Shimomura1999-nt, Iino2009-al}.
In the next section, we explore the functional role of a two-state model for goal-directed navigation.

\subsection{Sensory-driven state transitions enables turning towards the goal direction}
\begin{figure}[!t]
\centering
\includegraphics[width=.8\textwidth]{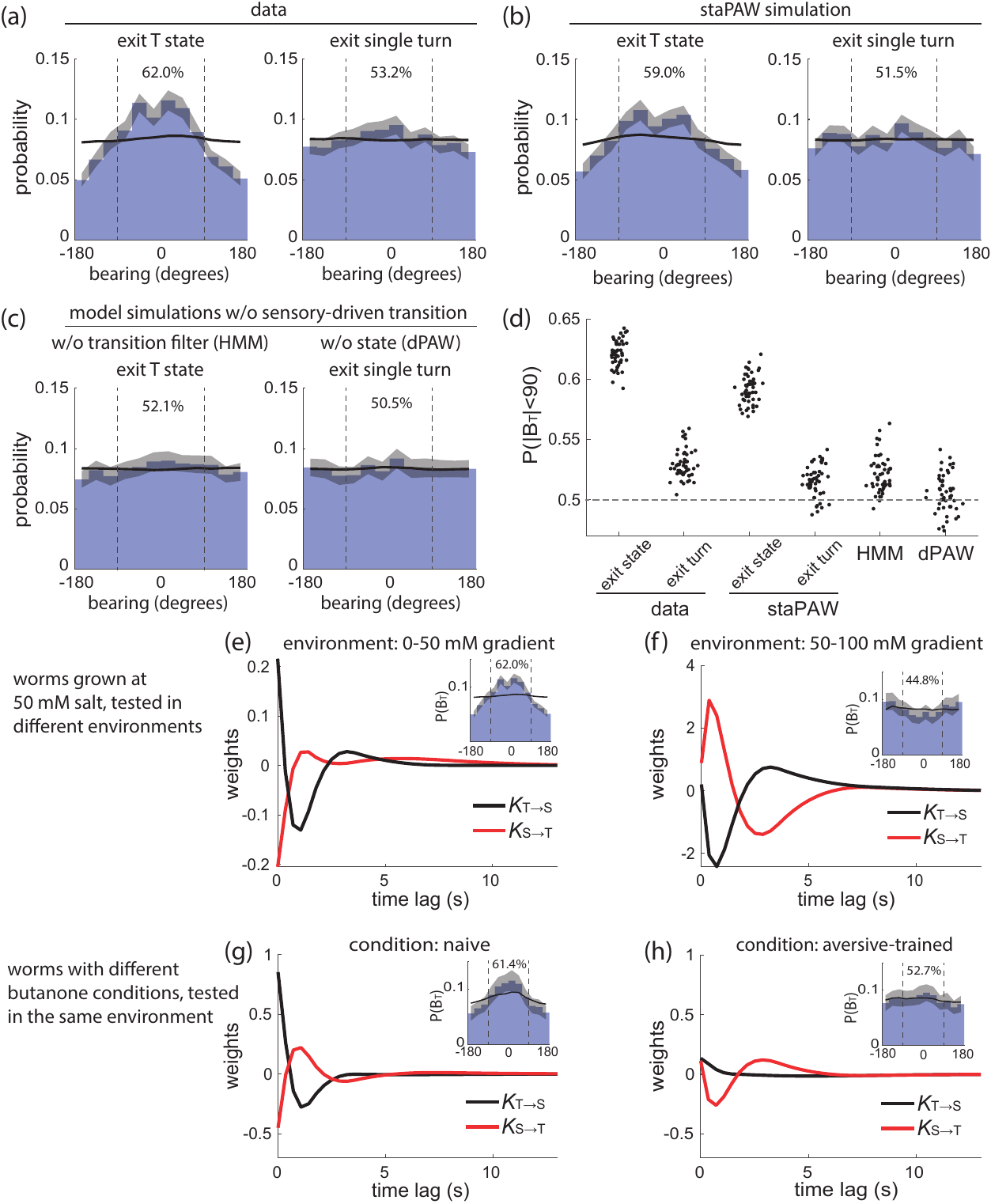}
\caption{
\textbf{Model state transition kernels govern how worms exit turn states towards the goal direction.}
\textbf{(a)} Distributions of bearing exiting the T-state (left) or a single turn (right) in worm navigation data.
\textbf{(b)} Same as (a) but for staPAW simulation. 
\textbf{(c)} Bearing distribution from models without sensory-driven states, including HMM and dPAW.  
\textbf{(d)} The probability of aligning to goal, when the bearing exit T-state $B_T$ (aligned when $|B_T|<90$ degrees) for data and different models. The scattered points show 50 repeats samples of 2000 events from the navigation trajectories. 
\textbf{(e,f)} Worms grown at 50 mM salt were tested in linear gradients ranging from (e) 0-50 mM or (f) 50-100 mM. Note that the kernels have opposite signs in two environments. The distribution of $B_T$ and the fraction aligned with gradient are shown in the inset. 
\textbf{(g,h)} Worms with different prior learning experiences were tested in the same butanone odor landscape. Naive worms (g) and aversive-trained (h) worms have altered kernels. 5-9 plates, each with $\sim$100 worms, are measured per condition.
}
  \label{fig:fig3}
\end{figure}

The mechanism that enables worms to exit a pirouette towards a favorable direction is not fully understood and is currently a source of active debate \cite{Pierce-Shimomura1999-nt, Tanimoto2017-pt, kramer2024neural}. The favorable exit could arise from a mechanism within individual turns, or it could arise from the grouping of turns within the pirouette. In our framework, the turn-enriched state corresponds to a pirouette. Indeed, worms are much more likely to exit the T-state oriented towards up gradient (62\%). 
We therefore asked whether this is due to a single-turn mechanism or one that relies on grouping turns into the T-state.

Our measurements indicate that the grouping of turns into the T-state is the primary explanation of exiting when the direction is upgradient. By contrast, exiting a single turn has at most a very small upgradient bias (53\%). We therefore conclude that the T-state preferentially ends on turns that exit upgradient (Fig~\ref{fig:fig3}a).

Sensory-driven state-switching provides exactly a mechanism for preferentially exiting the T-state upon turns that end upgradient. The state transition kernels for $T \rightarrow S$ acts like a positive derivative on concentration (Fig~\ref{fig:fig1}h), and therefore when the animal is oriented upgradient it increases the probability of transitioning out of the T-state. Indeed, simulating new traces from staPAW recapitulates the observed bias in turn orientation on T-state exit (Fig~\ref{fig:fig3}b). In effect, by grouping turns together, the T-state samples different headings and preferentially exits the T-state when the animal orients in a favored direction.   
We argue that this is a form of active sensing, where the worm actively samples local concentration during the turn-enriched states to make a better estimation of the favorable navigation direction.

A two-state model that lacked sensory-driven state transitions (equivalent to an HMM) does not trivially exhibit a bias upgradient, nor does the single-state  dPAW model, as expected (Fig~\ref{fig:fig3}c). We therefore conclude that it is the sensory-driven transition between states, and not the mere grouping of turns together, that confers the animal with the ability to exit the T-state upgradient. 
The staPAW model that incorporates these sensory-driven state transitions best captures the directed bearing angle observed in measurements (Fig~\ref{fig:fig3}, SFig~\ref{fig:SI3}). 
These results explain quantitatively and mechanistically how pirouettes produce directed turns that correct the animal's course during navigation \cite{Pierce-Shimomura1999-nt}. The staPAW model not only recapitulates the directed turn property, but also provides the underlying computational mechanism with sensory-driven state transitions, which is absent in alternative descriptions without states.

The worm's sensory behavior can exhibit different types of navigational goals and is not limited to purely navigating upgradient. For instance, in salt chemotaxis, animals can also navigate down-gradient, depending on the initial conditions \cite{Luo2014-pc, Iino2009-al}. We wondered whether sensory-driven state-switching is a general feature of goal-directed navigation in \textit{C. elegans}. In particular, we wondered whether the state-switching model and its corresponding kernels would reflect features of the goal, even for goals that were up- or down-gradient for different sensory modalities.

We first fitted staPAW to measurements of worms undergoing bidirectional salt chemotaxis. Worms were cultivated at 50 mM salt concentration, and placed in one of two linear gradients, either 50-100 mM or 0-50 mM. Worms navigate towards the salt concentration that they were cultivated at, corresponding to down- or upgradient in the two environments.  If sensory-driven state-switching contributes to goal-directed navigation in both these contexts, we would expect the animal to preferentially exit the T-state in the direction towards the goal, and we would expect staPAW to extract state-switching kernels that have a sign corresponding to the direction of the goal (up- or down-gradient). Indeed, when exiting the T-state, animal's bearing was preferentially biased towards the goal regardless of whether the goal was up- or down-gradient (Fig~\ref{fig:fig3}e,f, insets). Moreover, the kernels for transitioning between T- and S-states for animals on the 50-100 mM gradient (Fig~\ref{fig:fig3}f) had opposite signs and a different scale but were otherwise similar to the corresponding kernels for animals on the 0-50 mM gradient (Fig~\ref{fig:fig3}e).
The result indicates that the worms alter their sensory-driven state transitions according to the preference of salt concentration.

We further investigated whether kernels of a state-switching model reflected learned goal-directed navigation for the volatile odor butanone (Fig~\ref{fig:fig3}g,h). Naive animals are mildly attracted to butanone while aversive-trained animals are much less attracted \cite{Cho2016-is, Chen2024-od}.  We applied staPAW to previously acquired butanone navigation trajectories \cite{Chen2024-od}. Bearing angles exiting the T-state were biased upgradient for naive animals but the bias was diminished or went away for aversive-trained animals. Similarly, the sensory-driven state-switching kernels for naive and aversive-trained animals differ, and the $K_{T\rightarrow S}$ kernel loses its biphasic nature after aversive-training (Fig~\ref{fig:fig3}h). This shows that sensory-driven state-switching is informative when applied to a different sensory modality, and that its kernels reflect changes to both the sign and magnitude of the goal.

Together, the state-switching strategy reflects the navigation goal by tuning the state transition kernels. State-switching behavior is not specific to climbing salt gradients, but extends to a volatile odor; and it reflects both the sign and magnitude of the goal.

\subsection{Sensory neuron activity causally drives state transitions}

\begin{figure}[!t]
\centering
\includegraphics[width=.63\textwidth]{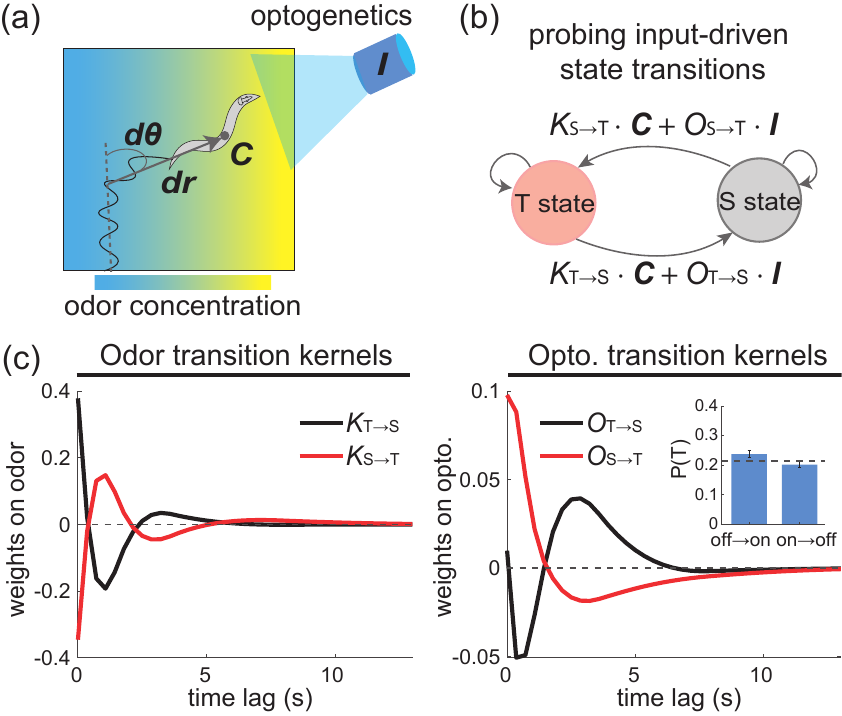}

\caption{
\textbf{Optogenetic stimulation shows that sensory processing causally drives state transitions.} 
\textbf{(a)} Experiment with concurrent presentation of odor $C$ and optogenetic input $I$. \textbf{(b)} Schematic for the staPAW model extended to incorporate optogenetic input that drive state transition probability. 
\textbf{(c)} State transition kernels for worms expressing ChR2 in AWC$^{\mathrm{ON}}$ sensory neuron. The butanone odor kernels $K_{i \rightarrow j}$ and optogenetic kernels $O_{i \rightarrow j}$ are shown. 
In the optogenetic case, inset shows the probability of being in the T-state for the 5s window when light is on versus a 5s window after light is off. Dash line shows the baseline probably before stimuli and error bars show standard deviation across 50 ensembles of 500 stimuli. T-test for transitions induced by stimuli show significant difference ($p<0.001$). The staPAW model that includes optogenetic input is fitted to data with over 3,000 pulses delivered during worm chemotaxis. 
}
  \label{fig:fig4}
\end{figure}

To pinpoint the sensory processing and neural circuits involved in state-switching navigation, we studied butanone navigation concurrently with optogenetic perturbation. Previously, we expressed ChR2 in a sensory neuron AWC$^{\mathrm{ON}}$ that is known to respond to volatile odorants \cite{Cho2016-is} and optogenetically stimulated it during the worm's navigation in a butanone odor landscape \cite{Chen2024-od} (Fig~\ref{fig:fig4}a). We now applied staPAW to this measurement and extended the model to include kernels that act on optogenetic stimuli in the state transition probability (Fig~\ref{fig:fig4}b).  Because AWC$^{\mathrm{ON}}$ is known to depolarize when the butanone odor is off \cite{Cho2016-is, Chen2024-od} we expect the animal to perceive optogenetic stimulation of AWC$^{\mathrm{ON}}$ as leaving butanone, and therefore it should drive it to the T-state. Indeed, we found that optogenetic stimulation of AWC$^{\mathrm{ON}}$ increases the likelihood of transitioning into the T-state (SFig~\ref{fig:SI4}a).
Similarly, we also expect and observe that state transition kernels for AWC$^{\mathrm{ON}}$'s activity resemble the sign-flip of the kernels for butanone stimuli (Fig~\ref{fig:fig4}c). This shows that the sensory-driven state transition kernels describe a causal relationship between sensory neuron activity and behavior state. Namely, optogenetically induced activity in AWC$^{\mathrm{ON}}$ increases the probability that animals transitions into T-state.

\begin{figure}[!t]
\centering
\includegraphics[width=.9\textwidth]{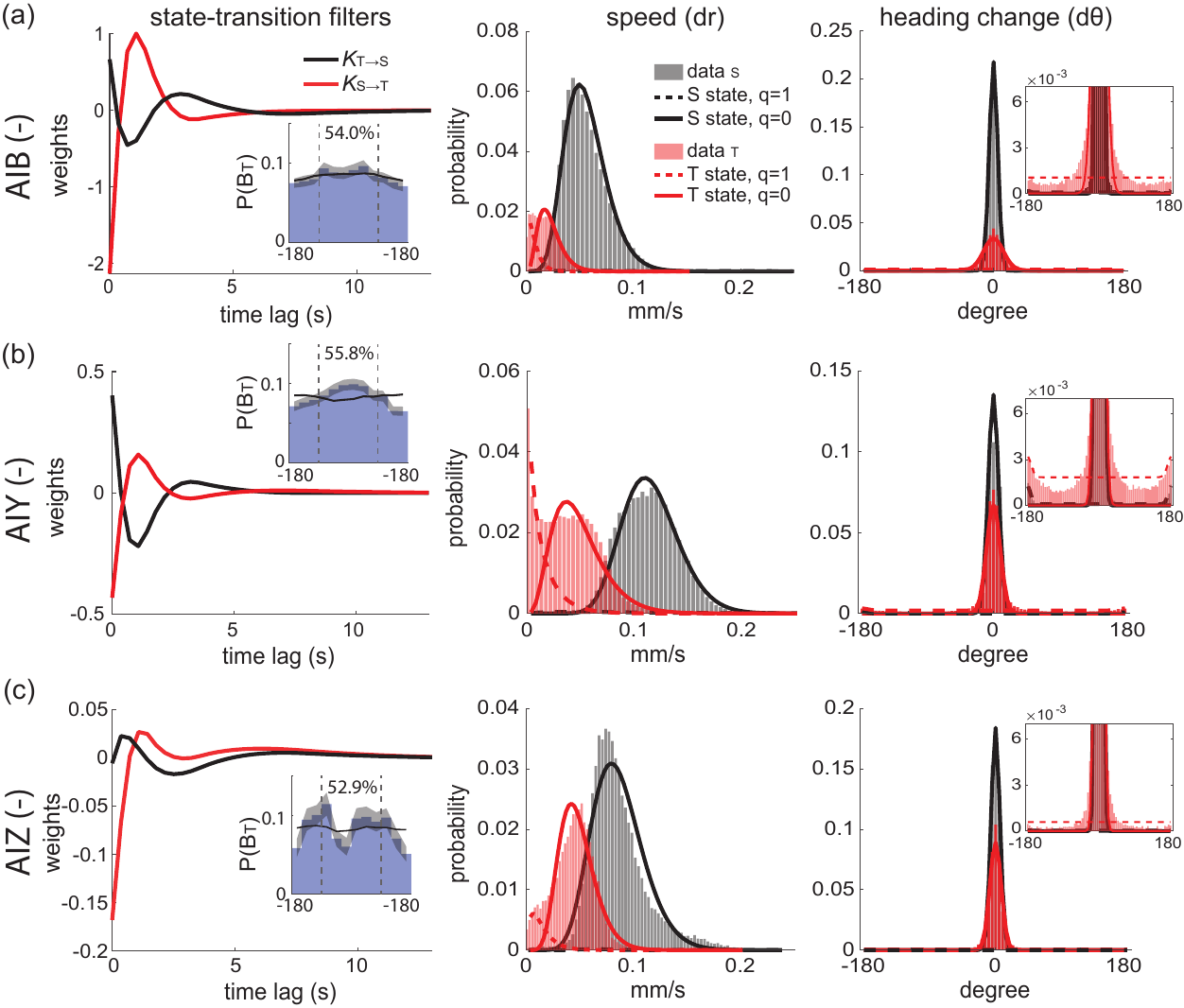}
\caption{\textbf{Disruptions to interneurons alter kinematics but not the state-switching dependence on sensory stimuli}
\textbf{(a)} State transition kernels fitted to worms with disrupted AIB navigating in a butanone gradient (left). 
The distribution of bearing exiting T-state ($P(B_T)$) and percentage of angles aligned to the goal are shown in the inset. 
The corresponding distribution for speed (middle) and heading change (right) of two states are shown. The enlarged probability for broad heading change is in the inset. 
\textbf{(b)} Same as (a) but for worms with disrupted AIY neuron. \textbf{(c)} Same as (a) but for worms with disrupted AIZ neuron. Each transgenic condition includes 5-10 animal hours of navigation trajectories for staPAW fitting.
}
  \label{fig:fig5}
\end{figure}

\subsection{Neural circuit control of sensory-driven state-switching is separate and distinct from control of kinematics}

We sought to gain insight into where in the circuit the state-switching control might reside. Specifically, we investigated contributions of neurons AIB, AIY, or AIZ because these three interneurons are direct downstream synaptic partners to both AWC and ASE neurons that sense odor and salt, respectively. Moreover, AIB, AIY and AIZ are known to be involved in chemotaxis behavior \cite{Chalasani2010-cp, Iino2009-al, Luo2014-pc}
. For example, AIB is involved in reversal behavior, AIY promotes forward movements during navigation and steering \cite{Chalasani2010-cp, gray2005circuit}, and  AIZ is needed for chemotaxis \cite{Iino2009-al, Chen2024-od}. We applied staPAW to measurements we had previously acquired of animals with genetic ablations in either AIB, AIY or AIZ undergoing naive butanone chemotaxis \cite{Chen2023-fy}.

Interestingly, neural perturbations dramatically changed the kinematics of movement within a state but did not change the fundamental rules that governed state-switching. To investigate changes to kinematics caused by neural perturbations, we inspected the distributions of the animal's speed and heading reflect. In the S-state, AIB-disrupted animals moved at half the speed of AIY-disrupted animals (Fig~\ref{fig:fig5}a,b, middle). In the T-state, AIZ-disrupted animals had almost no turns, while AIY-disrupted animals had many turns (Fig~\ref{fig:fig5}b,c, right). 

Despite the significant differences in the kinematics, at the level of state transitions, the overall rules governing their dependence on sensory stimuli were mostly preserved (Fig~\ref{fig:fig5}, left). While the amplitudes of the state transition kernels varied with interneuron disruption, the flipped-sign kernel that drives transitions between the S and T-states was preserved for genetically ablations of AIY and AIB, and only slightly phase-shifted in $K_{T\rightarrow S}$ for AIZ. This means that the neural perturbation to AIY, AIZ and AIB cause different rates of transitions between the S- and T-states, but their dependence on sensory stimuli is largely preserved. The chemotaxis performance of these mutant strains is reported in (SFig~\ref{fig:SI_BR}). Notably, staPAW learns parameters fitted to each strain that, when combined to form an effective drift velocity  upgradient, are  predictive of the measured chemotaxis index (SFig~\ref{fig:SI_BR}). This provides further evidence that staPAW captures chemotaxis even in  mutant strains with altered kinetics. 
More broadly, these results suggest that neither of the three interneurons alone, at least, are responsible for driving state transitions in response to sensory stimuli. 
Furthermore, it suggests that the neural mechanism of behavioral kinematics may be distinct from those that control sensory-driven transitions between states which is consistent with long-standing ideas of hierarchical control of behavior \cite{tinbergen1963aims, berman2016predictability, kaplan2020nested}.

\subsection{Sensory-driven state-switching is beneficial for chemotaxis and emerges from data-constrained optimization} 
 
\begin{figure}[!t]
\centering
\includegraphics[
width=.8\textwidth]{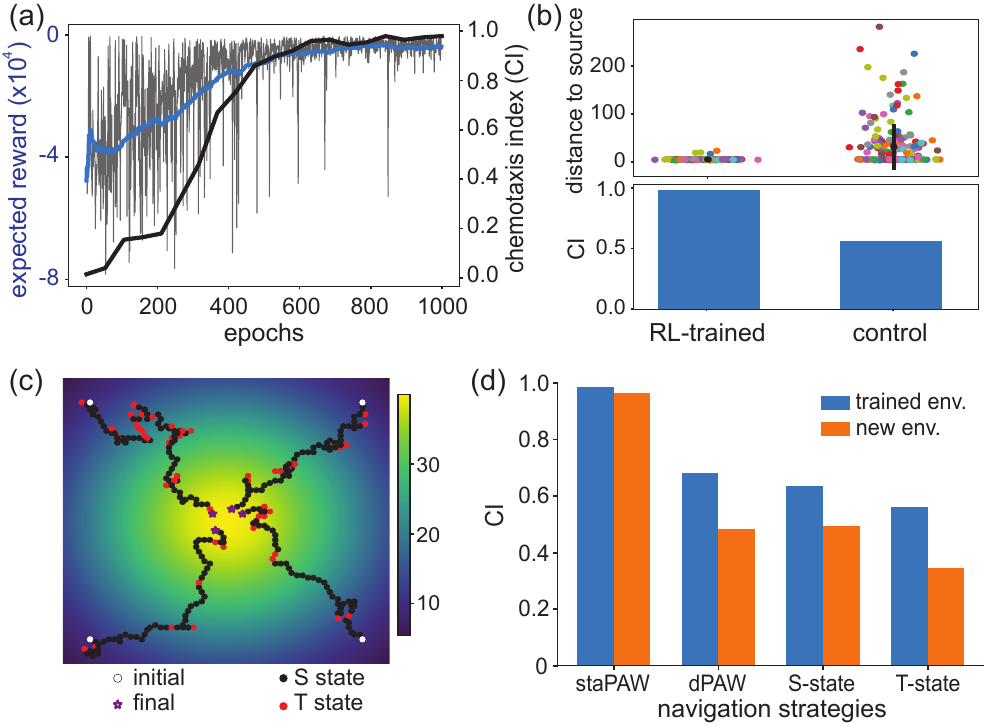}
\caption{\textbf{Data-constrained reinforcement learning model suggests that state-switching is beneficial for navigation performance.} \textbf{(a)} Expected reward and chemotaxis index (CI) as a function of epochs during reinforcement learning.  CI is the fraction of tracks that end up near the source. 
The blue line is a running average of expected rewards of every 50 epochs and the black line is the CI readout every 50 epochs. \textbf{(b)} Agent's distance to the point source at the end of navigation (top) and CI (bottom) from 200 trajectories in the RL-trained model compared to a control agent with sensory independent state transitions. The mean and standard deviation for distance are shown in black.
\textbf{(c)} Four example navigation tracks with the RL trained parameters. Two states are colored in black and red. The initial position is in white and the final position is in purple stars.  
\textbf{(d)} CI across different navigation strategies in two environments. 
Results in the trained environment with a Gaussian spatial profile and a new environment with shallower gradients (1.7 times the Gaussian spatial width).
}
  \label{fig:fig6}
\end{figure}

To quantitatively understand whether and how sensory-driven state transitions benefit chemotaxis performance, we conducted an \textit{in silico} investigation using a simpler model constrained to data. This model also has two states, but it \textit{a priori} doesn't yet know how to drive state transitions based on sensory stimuli.  The model is instead forced to learn, via reinforcement learning (RL), an optimal sensory dependence for state-switching to achieve efficient chemotaxis.  Within each S- and T-state, the agent undergoes sensory-driven weathervaning or pirouettes, but its kinematics are constrained to follow measured heading statistics observed in Fig~\ref{fig:fig2}b or c, respectively. This data constraint differentiates the agent from typical RL models used in the past that only take cardinal actions on a grid \cite{verano2023olfactory, Kim2020-sd, Yamaguchi2018-rh}. To keep the number of free parameters low, we also restrict the state-switching kernels to extend only one time step into the past. We placed agents in a simulated chemical landscape where they then learned to climb up gradient by optimizing the sensory dependence of their state-switching via reinforcement learning.

By adjusting only the sensory-dependence of their state-switching, the RL-trained agents learned to perform chemotaxis (Fig~\ref{fig:fig6}a,b) and generated trajectories with state transition behaviors (Fig~\ref{fig:fig6}c) that were qualitatively similar to measurements from animals (Fig~\ref{fig:fig1}g). 
Sensory-dependency of state-switching was initiated randomly, and agents at first failed to perform chemotaxis, but through RL optimization they eventually learned to achieve high performance (Fig~\ref{fig:fig6}a).

The sensory dependence on state-switching was crucial for achieving this performance. By contrast, control animals that were forced to transition between states in a sensory-independent manner failed to consistently reach the source and had lower chemotaxis index (Fig~\ref{fig:fig6}b).

The increase in chemotaxis performance from RL training is purely due to state-switching. The kinematics both within a state and in aggregate are the same for RL trained agents and for untrained control agents with enforced sensory independent state transitions (SFig~\ref{fig:SI_RL}a). 
Moreover, without state-switching, agents cannot perform nearly as well. Agents that are forced to reside in one state (S- or T-state), or that have no concept of state (dPAW) all fail to perform as well (Fig~\ref{fig:fig6}d).

Interestingly, the empirically observed behavioral  dynamics (Fig~\ref{fig:fig0}) that had initially suggested the presence of alternating persistent states emerge naturally from RL-training (SFig~\ref{fig:SI_RL}b,c).  The two distinct timescales in the distribution of inter-turn intervals (Fig~\ref{fig:fig0}c) emerges from  RL-training and is not present in untrained control animals (SFig~\ref{fig:SI_RL}b). The ability to preferentially exit T-states in the upgradient orientation (Fig~\ref{fig:fig0}f) is also dramatically enhanced by RL-training   (SFig~\ref{fig:SI_RL}b).  
We therefore conclude that state-switching is beneficial for navigation because it is an optimized solution in a reinforcing learning paradigm. Moreover, features like the directed exit from the T-state emerge naturally from optimization.

\begin{figure}[!t]
\centering
\includegraphics[
width=.66\textwidth]{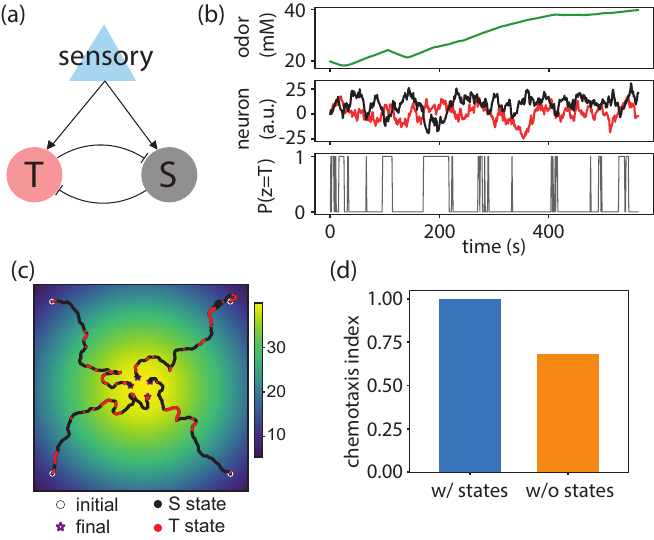}
\caption{
\textbf{An effective circuit model recapitulates state-dependent navigation.} \textbf{
(a)} Circuit model with mutual inhibiting turn- and steer-enriching neurons. \textbf{(b)} Example time series of concentration input (top), the neural activity of two units (middle), and the state assigned to turn-enriched state (bottom). \textbf{(c)} Example time series of navigation trajectories (upper right trace corresponds to (b). \textbf{(d)} Chemotaxis index computed from simulations of the models with and without the state-dependent circuit motif.
}
  \label{fig:fig7}
\end{figure}

\subsection{A minimal mutual inhibition circuit motif is sufficient to generate sensory-driven state-switching}

The staPAW model and the data-constrained RL model are both phenomenological descriptions of the worm's navigation behavior. We further explored a biologically inspired minimal neural circuit model that captures state-switching navigation strategies (Fig~\ref{fig:fig7}a). 
Our effective minimal network model consists of two mutually inhibiting rate neurons that both receive sensory input. The two neurons correspond to two different navigation states, and the model produces a neuron-dependent behavioral strategy according to the neuron with larger activity. 
This simple circuit model recapitulated similar state-switching navigation time series (Fig~\ref{fig:fig7}b,c). Furthermore, we compared the chemotaxis performance of agents in a simulated odor environment with and without the mutual inhibiting motif that generates state-dependency. The minimal circuit model is consistent with observation in ((Fig~\ref{fig:fig7}d), where models with state-switching have higher chemotaxis compared to models without states. 

This minimal circuit model suggests that one way of generating state-switching behaving would be to have mutual inhibition between two neurons or sets of neurons, which, if ablated, would destroy state-switching. Notably, if such a motif exists in the worm, results from genetic ablation studies shown in Fig~\ref{fig:fig5} suggest that neither AIB, AIY and AIZ alone, constitute an inhibitory node. The mutual inhibitory neural circuit can motivate future searches for neural correlates of T- or S-states.

\section{Discussion}

We investigated how worms produce persistent behavior and directed turns during sensory navigation. 
Our model in which the animal has states that persist in time better describes experimental measurements compared to models without state-dependency. The animal exhibits two states-- one state that elicits high turning probability (T-state) and another state that is enriched for runs (S-state). Each internal state describes a set of behavior statistics and sensorimotor input-output relations that persist in time. 
Through optogenetic perturbation, we verified that sensory input causally drives the animal to transition between these two persistent states. This sensory-driven state-switching enables worms to turn towards the goal direction. 
We present evidence to suggest that the neural basis of sensory-driven state-switching is separate and distinct from the neural basis of the animal's detailed kinematics, and that state-switching is beneficial for the worm's navigation.

The two states revealed by staPAW  are similar to pirouettes and runs as originally defined by \cite{Pierce-Shimomura1999-nt} --- in that work ``pirouettes'' refer to bouts of turns and runs refer to everything else. In our current framework, the definition of a S- and T-state emerges naturally in an unsupervised manner by fitting a sensory-driven multi-state model to the experimental measurements. Indeed, the two-state nature (and not three, or four) emerges from the data and the same procedure is adaptable to genetically perturbed animals that have different behavioral statistics (Fig~\ref{fig:fig5}). By contrast, in the past identifying pirouettes required an \textit{a priori} chosen threshold to define a turn, and then additional hand-tuned thresholds were applied to the inter-turn intervals \cite{Pierce-Shimomura1999-nt}. Other approaches also required hand tuned parameters or pre-defined time-windows \cite{Iino2009-al, Tanimoto2017-pt, soh2018computational, Luo2014-pc, Roberts2016-rs}, or lack persistent state-dependency entirely \cite{Dahlberg2020-ip, appleby2012model, dunn2007circuit}. More recent methods apply delay-embedding to capture the continuous worm behavior \cite{Ahamed2021-wi, Costa2023-yv}, revealing the consistent pirouette state. But none of these approaches reveal the explicit sensory dependency of transitions both into and out of the T-state or pirouette. Here our definition of states requires no hand-tuned parameters, and we explicitly test alternative hypothesis for the organization of states such as the number of states, and how they depend on sensory stimuli.

A key finding of this work is that the S- and T-state transitions occur on few seconds timescales, are sensory-driven, and are relevant and beneficial for goal-directed navigation. These attributes distinguish it from other locomotor states that have been described in the literature. For example, the timescale of the S-state is much longer than an individual turn, but much shorter than the minutes-long timescale  of ``roaming'' or ''dwelling''--- states associated with finding sparse food patches \cite{Flavell2013-pl, kaplan2020nested, Bartumeus2016-kw, margolis2024stochastic}. The existence of different states on different timescales is an example of hierarchy in the organization of behavior\cite{tinbergen1963aims, berman2016predictability}.

Importantly, we show that alternating between S- and T-states not only better captures the animal's behavior, but it quantitatively improves gradient climbing, suggesting an integral role in chemotaxis. While many prior works have been very successful in identifying states that capture animal behaviors \cite{Calhoun2019-ci, berman2016predictability, datta2019computational, liu2018temporal, Wiltschko2015-uq}, it has often been challenging to connect those states to clear goals or functions.  A unique contribution of this work is that it explicitly connects states to goals.

The sensory-driven state transitions resolve an apparent mystery of how the worm is able to preferentially exit pirouettes upgradient \cite{
Pierce-Shimomura1999-nt, Iino2009-al, Tanimoto2017-pt, kramer2024neural}. This is currently an active area of investigation, with concurrent work seeking the neural basis for exiting turns upgradients\cite{kramer2024neural}. 
Here we make no claim about the worm's ability to exit single isolated turns upgradient, which would seemingly require the worm to keep in its head both the gradient and its moment-to-moment heading throughout a turn-- a capability that had not previously been demonstrated. We instead find evidence that the worm exits the T-state (similar to pirouettes) upgradient \cite{Pierce-Shimomura1999-nt} (Fig~\ref{fig:fig3}a). The sensory-driven nature of switching from T- to S-states provides a simple explanation for how this occurs: the worm resides in the T-state and continues to make a series of turns until it is oriented upgradient at which point the sensory stimuli drive it to transition into the S-state (Fig~\ref{fig:fig1}h). This explanation does not require the animal to measure or remember the gradient during the turn or to maintain any moment-by-moment knowledge of its heading during the turn.

Neural control of state transitions appears to be distinct from the neural control of the worm's detailed movements or kinematics because genetic perturbations to first-layer interneurons resulting in altered kinematics does not seem to dramatically alter sensory-driven state transitions. This implies that state-dependent control may be deeper in the network or may be implemented by sets neurons of and not any individual. Our toy model suggests that one way the nervous system could potentially implement state-switching is through mutual inhibition circuit motifs \cite{Roberts2016-rs, meng2024tonically, ji2021neural}. Future work could look for these motifs or other mechanisms of state control. For example, neurons implicated in exiting turns upgradient may also have a role in mediating sensory-driven T- to S-state transitions \cite{kramer2024neural}. Population-level neural imaging in behaving animals can reveal neural dynamics underlying behavioral state transitions\cite{Nguyen2016-ey, Atanas2023-bz}.

Our model is a variant of a GLM-HMM model, and the specific design choices of the model enabled many of these scientific discoveries. 
GLM-HMMs are one of a general classes of state-space models that have been widely used to characterize time-varying animal behavior \cite{Wiltschko2015-uq, Linderman2019-mj, datta2019computational, Calhoun2019-ci, Ashwood2022-he, bolkan2022opponent, Buchanan_undated-go, Reddy2022-qi}.
Most prior GLM models describe discrete behavioral choices across trials \cite{Ashwood2022-he, bolkan2022opponent, Calhoun2019-ci},
but here we have extended this framework to continuous time series. The extension to continuous time series is important because it allows the model to capture how the time-varying odor signals drive the time-varying behavioral kinematics that are essential for chemotaxis.
We have also chosen to use an input-driven HMM to model state transitions \cite{Calhoun2019-ci}. By using an input-driven HMM the model is capable of capturing the sensory-driven state transition which we show are relevant beneficial for gradient climbing.

Note, here by using an HMM we fit a Markov state transition model to time series data, but we are not claiming that the worm's behavior is itself Markovian. Conceptually, the model can be continuously driven by sensory inputs, which breaks the Markovian nature. More directly, we observed that the animal's state occupancy slowly drifts on the many minutes timescale (SFig~\ref{fig:SI_states}b), consistent with known longer-timescale processes, such as roaming and dwelling \cite{Gordus2015-kk, Flavell2013-pl}. Future extension to our modeling framework could explore these processes, for example by adding another modeling layer to the hierarchy shown in Fig~\ref{fig:fig1}a.

The GLM-HMM modeling framework provides ``bottom-up'' evidence for state-switching. By comparing to the detailed kinematics of animal behavior, the GLM-HMM shows how a two-state model better fits animal's observed behavior than the alternatives. But an important element of this work is the use of a reinforcement learning (RL) framework in a ``top-down'' way to provide independent evidence that sensory-driven state-switching is functional and beneficial. In the RL-framework it is the goal that is specified -- navigate upgradient.  RL is then used to learn an optimal strategy constrained only by the general statistics of the animal's movements. Using RL, a two-state sensory-driven state-switching strategy emerges naturally as an optimal way to navigate upgradient.

The use of general behavioral statistics to constrain the RL model is a critical advance. Prior uses of RL to describe animal behavior typically have agents that walk on a grid and lack constraints on their spontaneous motion \cite{Yamaguchi2018-rh,Ashwood2022-he, Kim2020-sd, Rigolli2022-ci}.  By contrast, here we constrain our RL agent to have the same distributions of movements as a real worm, which forces the RL model to be more realistic and allows us to make stronger claims related to experimental observations. 
We expect that this combination of state-space modeling and data-constrained RL will help reveal behavioral strategies and functions across species and tasks in future studies.

\section{Methods}

\subsection{Worm strains and preparation}

Worms were grown on nematode growth medium (NGM) plates, fed with OP50 bacterial food, and kept at 20C. Before chemotaxis experiments, we synchronized batches of worms and conducted experiments with day-1 adult worms. For optogenetic experiments, worms were kept on food containing 10 $\mu$l all-trans retinal. For genetic ablation, worms were treated with blue light stimuli at L1 stage. Details of the protocol are reported in \cite{Chen2024-od}.

For optogenetics, we used AML105 strain for AWC::ChR2 and QW910 strain for RIM::ChR2 experiments. For genetic disruption studies, we used AML580 for AIB down-regulation, IK2962 for AIY ablation, and IK3241 for AIZ ablation studies. These strains have been reported in prior work \cite{Chen2024-od, Chen2023-fy, ikeda2020context}.

\subsection{Chemotaxis experiments and behavioral imaging}

For butanone odor experiments, we used the instrument reported in \cite{Chen2023-fy} to control and measure odor concentration in a flow chamber. For salt chemotaxis experiments, we followed experimental method in \cite{Luo2014-pc} to make linear salt gradients ranging from 50-100 mM and 0-50 mM in a 9 cm square lid. The experiments were conducted in the same flow chamber and imaging system but without airflow during salt experiments. Measurements from naive and aversive-training in butanone chemotaxis and genetic ablation studies have been reported in \cite{Chen2024-od}. Salt chemotaxis and optogenetic perturbation experiments include new measurements in this work.

The behavioral imaging pipeline is modified from \cite{Chen2023-fy}. We tracked the center of mass of each worm in the arena at 14 Hz. The trajectories were smoothed and down-sampled to a data point every 5/14 second. The instantaneous speed and angle change were measured along the trajectory. We removed tracks that move less than 1cm$^2$ or tracked for less than 3 minutes. The measurements in this work includes $\sim$15-20 animal hour of recordings per navigation condition.

\subsection{The staPAW model and inference}

The staPAW model describes how the consecutive step speed $dr_{t+1}$ and heading change $d\theta_{t+1}$ are governed by experimental observations including tangential concentration $C_t$, perpendicular concentration difference $dC^{\perp}_t$, past behavior ($d\theta_t$ but not $dr_t$), and optogenetic input $I_t$ presented in the perturbation experiments. The emission of speed and heading change also depends on a latent state $z_t$ that is modeled as an input-driven Markov chain. Within a state, it is similar to dPAW model that makes the Bernoulli decision $q_t$ to perform a turn ($q=1$) or to weathervane ($q=0$). Similar to the proposed formulism in \cite{Chen2024-od}, the staPAW model conditioned on the latent state $z_t$ is written as shown in equation \ref{eq:2}, with the decision to turn formulated as:
\begin{equation}\label{eq:4}
    P(q_t=1  | z_t ) = m^z+ \frac{M^z-m^z}{1+\exp
(K_C^z \cdot \mathbf{C}_{1:t-1} + K_h^z \cdot |\mathbf{d\theta}_{1:t-1}|)} 
\end{equation}
which is the mixing probability of turning versus weathervaning behavioral strategies. Note that in the previous dPAW model \cite{Chen2024-od} the individual turns were called ``pirouettes,'' consistent with that model's lack of a notion of state or hierarchy. In the current hierarchical model we call this decision a single turn, and it is groupings of turns into the T-state that resembles the classical definition of pirouette \cite{Pierce-Shimomura1999-nt}. 
The state-dependent parameters are denoted with superscript $z$, with the maximum $M$, minimum turn probability $m$, turning kernel $K_C$, and history behavioral kernel $K_h$. We assumed that the conditional probability of speed and heading change is independent: $P(dr,d\theta)=P(dr)P(d\theta)$, for simplicity. The speed is modeled with Gamma distribution $P(dr|a,b)$ with shape $a$ and scale $b$ parameters conditioning on turn and weathervaning strategies $q$. The heading change densities follow:

\begin{align}\label{eq:56}
P_{\mathrm{tr}}(d\theta) &= \alpha^z U[-\pi, \pi] + (1-\alpha^z) f(\pi, \kappa_{\mathrm{tr}}^z), \\
P_{\mathrm{wv}}(d\theta \mid dC^{\perp}_{1:t-1})&= f(-K_{dC^{\perp}}^z \cdot \mathbf{dC}_{1:t-1}^{\perp}, \kappa_{\mathrm{wv}}^z)
\end{align}
where $\alpha$ is the mixture weight on uniform distribution $U$, $f$ is the von Mises distribution with mean and variance $\kappa$ parameters. The steering kernel $K_{dC^{\perp}}$ acts on the perpendicular concentration change to steer the heading during weathervaning.

In addition to emissions that are concentration dependent (equation \ref{eq:2}), the state transitions of $z$ are also input-driven, with transition probability following a soft-max form (equation \ref{eq:1}).


For optogenetic experiments, we added in terms $O_{ij} \cdot \textbf{I}_{1:t-1}$ in the state transition probability (equation \ref{eq:1}) and $K_O \cdot \textbf{I}_{1:t-1}$ in the turn decision function (equation \ref{eq:2}). The state transition kernels and turning kernels for optogenetic input are denoted as $O$ and $K_O$, respectively.

For inference, the experimental data for fitting is $D = \{ C, dC^{\perp}, d\theta, dr, I \}$ and the history components with time lag is denoted $H$. 
The objective for fitting staPAW is the log-likelihood that sums over latent states. The joint distribution of data and latent is the complete-data log-likelihood with all parameters $\Theta$:

\begin{equation}\label{eq:7}
    \text{ECLL}(\Theta) = \Sigma_z P(z|D, \Theta^{\text{old}}) \log ( P(dr,d\theta,z | H,\Theta))
\end{equation}

We optimized this objective through expectation-maximization (EM) algorithm. The E-step applies forward-backward algorithm to estimate the marginal and the joint posterior state probabilities and the M-step uses the state estimates to compute and maximize the $\text{ECLL}$. In practice, we initialized parameters in the EM runs around the maximum likelihood estimate of a single state (dPAW model). We repeated different insinuations for 5 times for each of the 3-fold cross-validation training set and choose the parameter with the highest training likelihood for testing in Fig~\ref{fig:fig1}. Due to the inclusion of strictly-positive parameters, such as turn probabilities, variance, and shape parameters, we performed constrained optimization during the M-steps. We empirically found that EM converges around 20-30 iterations.


\subsection{Reinforcement learning model}

The RL model constrained by worm navigation takes actions that correspond to the behavioral states fitted in staPAW. Here we simplified the model by using constant speed and state-dependent heading changes. The policy takes in a local concentration difference measurement $\Delta C(s)$ at location $s$ across a single time step and makes the decision following:

\begin{equation} \label{eq:8}
    \pi_{\theta}(a=1|\Delta C(s)) = \frac{1}{1+\exp(K \Delta C(s) + b)}
\end{equation}
where $K$ is the scalar weight on concentration difference $\Delta C(s)$ and $b$ is a baseline. These are the model parameters $\theta$. This policy is similar to the soft-max state transition in staPAW (equation \ref{eq:1}) but simplified for concentration difference in a single step. Given the action that corresponds to state-dependent kinematics (speed and heading change), the worm navigates to the next location in the environment $P(s'|s,a,\Delta C)$.  This model setting forms a partially observed Markov decision process (POMDP), since the policy only takes in a local concentration difference in the environment, rather than the actual states in the environment. The POMDP setting is similar to formulation in previous navigation RL models \cite{verano2023olfactory, Rigolli2022-ci}.

The reward is a function of the state and action, where state is the coordinates in the environment $s$ and action is the discrete output from equation \ref{eq:8}. The function is a combination of the negative distance to source $s^*$ and a filtered action $\phi(a)$ weighted by scalar $\lambda$:
\begin{equation} \label{eq:9}
    r(s,a) = -d(s,s^*) - \lambda\phi(a)
\end{equation}
where the function $\phi$ is an exponential filter on the discrete actions to penalize behavioral switching. This term is added to qualitatively recapitulate the experimentally observed state occupancy of behavioral states.

The parameters in the policy was optimized through policy gradient method (Fig~\ref{fig:fig6}): 
\begin{align}  \label{eq:1011}
    \theta' \leftarrow
 \theta + \eta \nabla J , \\
    J(\theta) = \mathbb{E}_{\pi_{\theta}} [\Sigma_{t=1}^{T} \gamma^t r^t]
\end{align}
where $\eta$ is the learning rate and $\gamma$ is the temporal discount factor used to compute the expected total reward across time $J$. In practice, we used Adam optimizer to adjust the learning rate during gradient descent and run for 1000 epochs and observed convergence of the expected reward. In the same model structure, we shuffled actions to make the dPAW model, and fixed the action at 0 or 1 for weathervaning or turn-only models, respectively.

\subsection{Neural circuit model}

The effective circuit model has neural activity following:
\begin{equation} \label{eq:12}
    \tau\frac{dx_i}{dt} = -x_i + \Sigma_j J_{ij}\phi(x_i) + B_i I_i(t) + \xi_i
\end{equation}
where $x_i$ is the voltage of neuron $i$, with time constant $\tau$, circuit connection $J_{ij}$, input weight $B_i$, and time-varying white noise $\xi_i$. Following a sensory response model for AWC neuron \cite{
molina2024conflict}, the input $I_i(t)=F(t)-S(t)$ is from the fast and slow signal processing concentration input $C(t)$ of the sensory environment experienced by the agent:
\begin{equation} \label{eq:13}
    \frac{dF}{dt} = \alpha C - \beta F,
\end{equation}
\begin{equation} \label{eq:14}
    \frac{dS}{dt} = \gamma(F-S)
\end{equation}
with coefficients $\alpha$, $\beta$, and $\gamma$ governing the concentration processing. In the effective circuit model, there are two units $i \in \{S, T\}$ corresponding to S and T-states (Fig~\ref{fig:fig7}a). Both units have the same parameters and only differ in the coupling strength $J_{ij}$. The agent employs behavioral strategy from the state that has a higher activity at any given time point. In practice, we simulated neural dynamics and navigation behavior with Euler method with $dt=0.1$ s and measured dwell time and chemotaxis index from 200 trajectories.

\subsection{Data and code}

All experimental setup and procedure are documented in previous work \cite{Chen2023-fy} and data uploaded \cite{Chen2024-od}. New measurements in this work and compiled data to make figures in this paper are available on FigShare: \url{10.6084/m9.figshare.25996498}. 
Code for staPAW model, analysis, and generating figures are available at \url{https://github.com/Kevin-Sean-Chen/Chemotaxis_function}. Code for RL training and simulations are available at \url{https://github.com/Kevin-Sean-Chen/wormRL}

\section*{Acknowledgments}
We thank the Leifer lab and Pillow lab for helpful discussions. We thank Gautam Reddy for helpful feedback on the staPAW model and reinforcement learning. 
We thank the Kavli Institute for Theoretical Physics at University of California Santa Barbara for hosting us during the completion of this work. We also acknowledge the Gordon and Betty Moore Foundation Grant No. 2919.02.
This work was supported by the National Institutes of Health National Institute of Neurological Disorders and Stroke under New Innovator award (DP2- NS116768), the Simons Foundation under award SCGB (543003), an NSF CAREER Award to AML (NSF PHY-1748958), the Kavli Institute for Neuroscience at Yale University (Kavli postdoctoral fellowship to KSC) and through the Center for the Physics of Biological Function (PHY-1734030).

\bibliographystyle{unsrt}  
\bibliography{worm_ssm}

\section*{Supplementary information}

\begin{suppfigure}[!h]
\centering
\includegraphics[width=.9\textwidth]{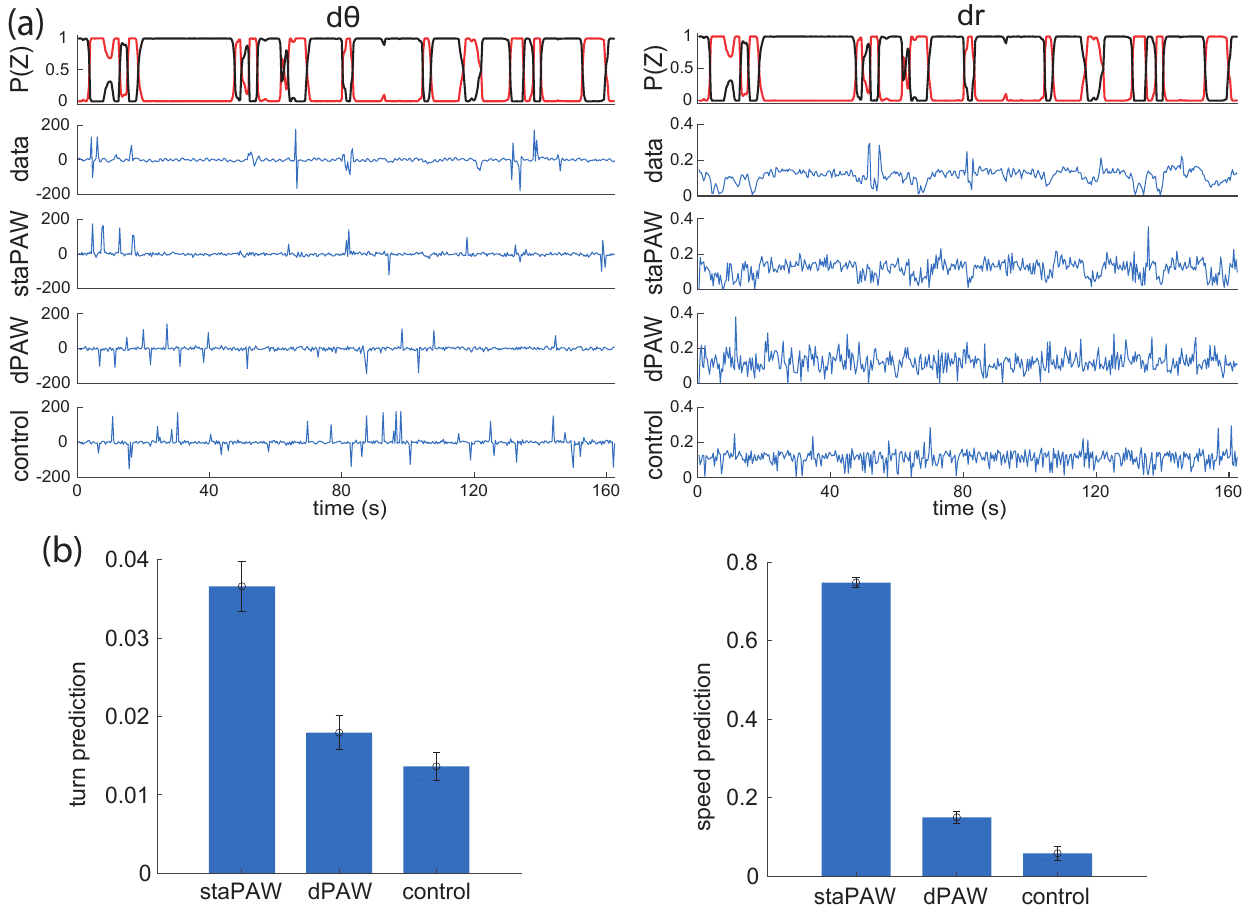}
\caption{The staPAW model better predicts time series of worm navigation behavior. 
\textbf{(a)} An example navigation time series with heading change (left) and speed (right) from experimental measurements (data) is shown in the second row. The posterior state probability and output of staPAW are shown in black (S-state) and red (T-state) lines. The output from dPAW and a control model without concentration input are shown below.\textbf{(b)} Across 500 sampled trajectories shown in (a), the prediction of turning is computed with Jaccard similarity index and prediction of speed is computed with correlation coefficient for three different models. Error bars show standard error of mean.
}
  \label{fig:SI1}
\end{suppfigure}

\begin{suppfigure}[!t]
\centering
\includegraphics[width=.9\textwidth]{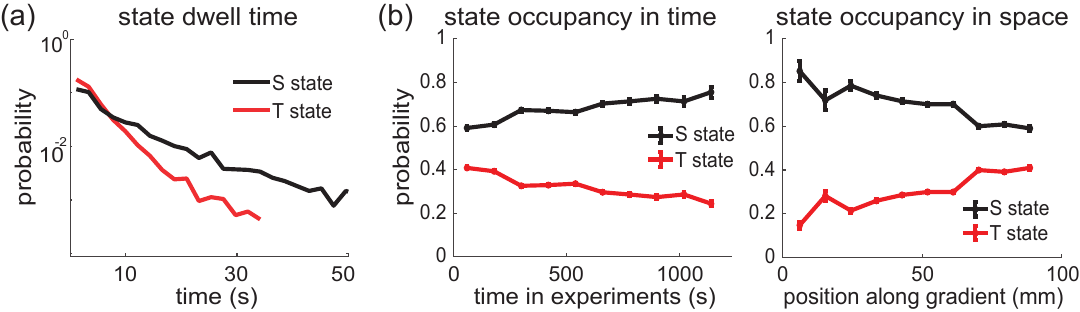}
\caption{Spatiotemporal distribution of states.  
\textbf{(a)} Dwell time distribution of two states in chemotaxis experiments. \textbf{(b)} The state occupancy across experimental time (left). Error bars show counting statistics of $\sim300$ navigation tracks. The occupancy along the position of the linear salt gradient is shown on the right.
}
  \label{fig:SI_states}
\end{suppfigure}

\begin{suppfigure}[!t]
\centering
\includegraphics[width=.9\textwidth]{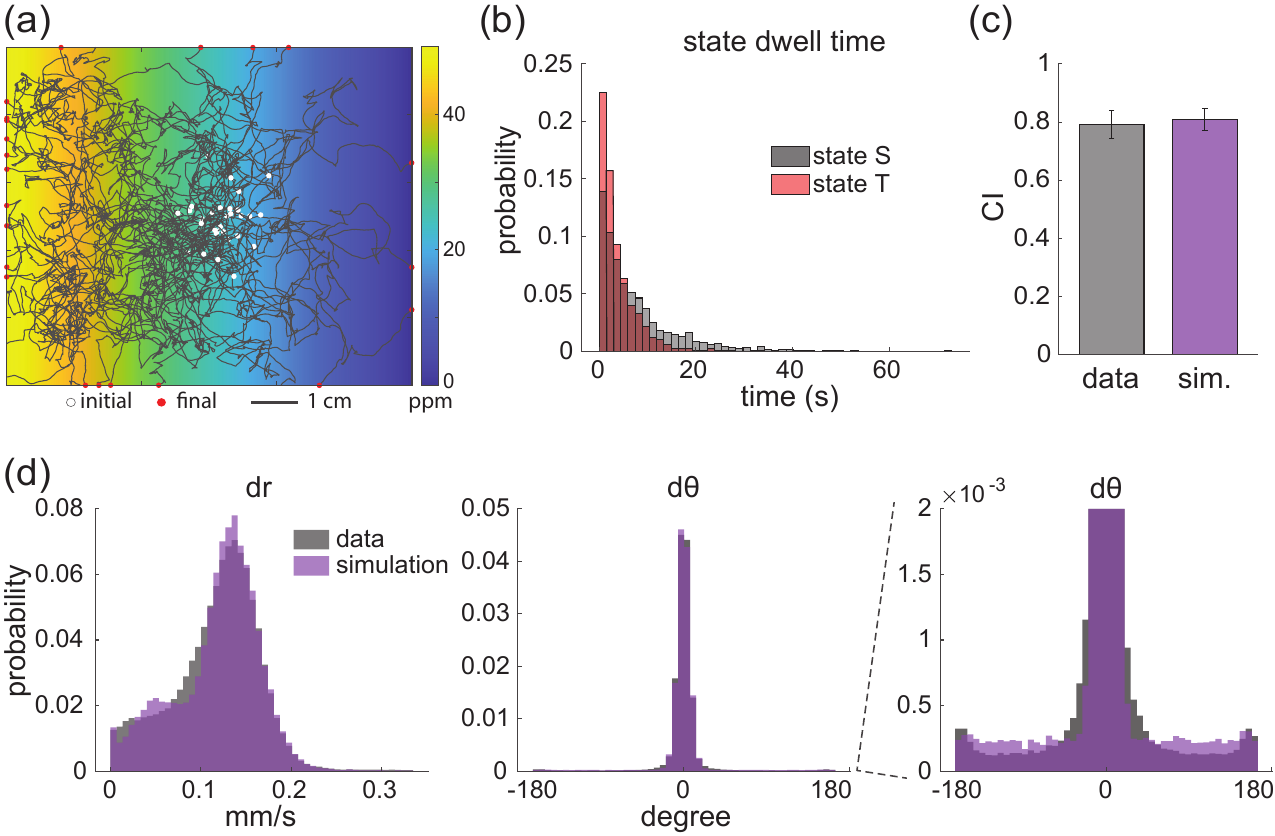}
\caption{The staPAW model generates navigation behavior similar to experimental observations. 
\textbf{(a)} Simulation of staPAW with parameters fitted to salt chemotaxis in the same environment. \textbf{(b)} The dwell time distribution of two states measured from the simulated trajectories in (a). \textbf{(c)} Chemotaxis index (CI) of the data and simulation. Error bars show standard deviation of 10 repeats of 50 sampled navigation trajectories. t-test shows no significant difference. 
\textbf{(d)} The distribution of speed $dr$ and heading change $d\theta$ for experimental data and simulations from the fitted staPAW model.
}
  \label{fig:SI2}
\end{suppfigure}

\begin{suppfigure}[!t]

\centering
\includegraphics[width=.8\textwidth]{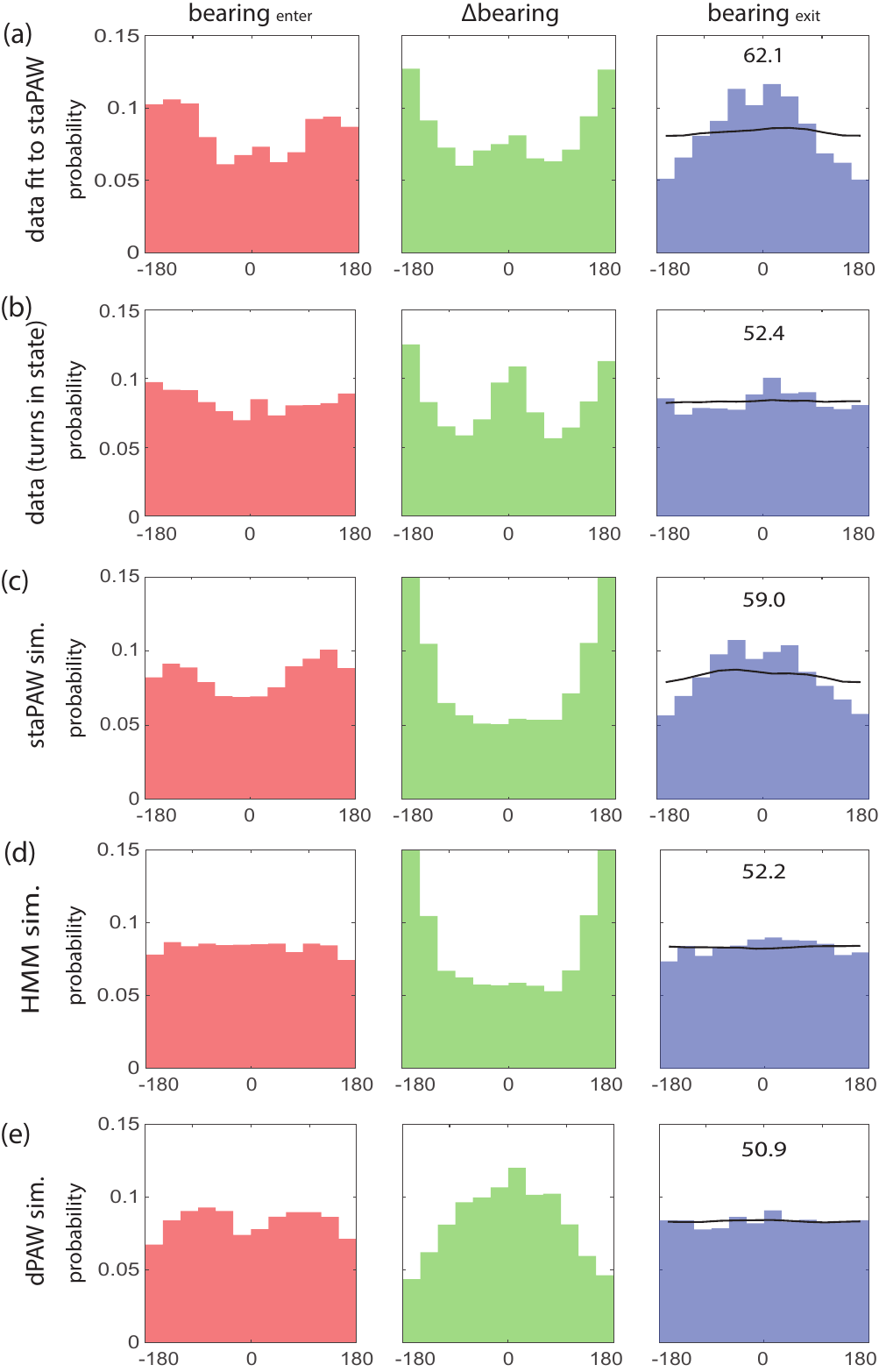}

\caption{Bearing angle around state transition for data and models. 
\textbf{(a)} The bearing angle before entering the T-state (left), 
the change in bearing angle in between (middle), and the bearing angle after exiting the turn (right). The bearing angle after shuffled $\Delta$bearing is shown in black line. 
The percentage of alignment to goal $P(|B_T|<90)$ while existing events are shown in the right panels. 
The histograms are computed for T-states inferred in data. 
\textbf{(b)} Same as (a) but for individual turns within a pirouette state.
\textbf{(c-e)} Same as (a) but for behavior simulated from models, including the full staPAW model (c), HMM that does not have sensory-driven states (d), and dPAW that does not have state-dependency (e).
}
  \label{fig:SI3}
\end{suppfigure}

\begin{suppfigure}[!t]
\centering
\includegraphics[width=.9\textwidth]{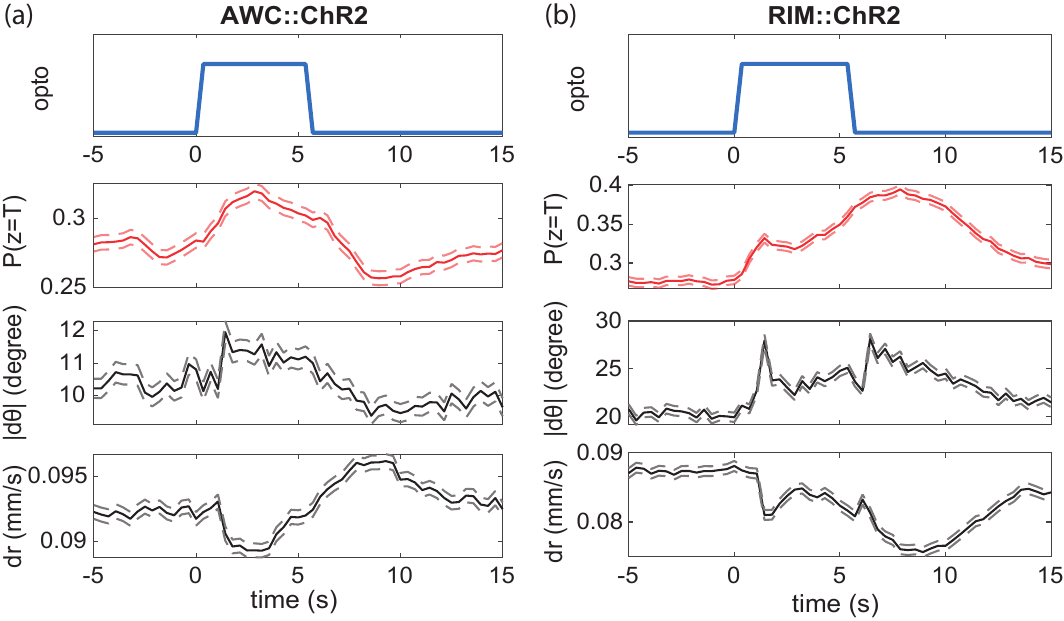}
\caption{The staPAW model captures state-dependent behavior driven by optogenetic input. 
\textbf{(a)} Behavioral response and the posterior probability of entering the turn-enriched state ($P(z=T)$) aligned to optogenetic input. The dash lines show standard error of mean. Measurements from 5000-6000 impulses are included. AWC::ChR2 strain is used for (a) and RIM::ChR2 is shown in \textbf{(b)}.
}
  \label{fig:SI4}
\end{suppfigure}

\begin{suppfigure}[!t]
\centering
\includegraphics[width=.6\textwidth]{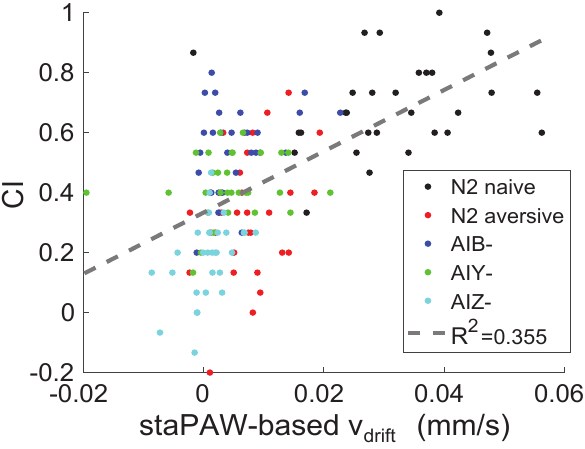}
\caption{Comparing model-predicted drift velocity ($v_{\text{drift}}$) and chemotaxis index (CI). Color code indicates different worm strains and conditions, each with 30 sub-sampled tracks, and dash line shows linear regression. The estimate drift is computed with $v_{\text{drift}} = \langle dr\rangle_S P(S)(P_{|B_T^{\text{exit}}|<90} - P_{|B_T^{\text{enter}}|<90})$, where $\langle dr\rangle_S$ is the average speed in S-state, $P(S)$ is the probability of being in S-state, and the last term is the difference between probability of aligned bearing exiting and entering T-state. This calculation simplifies the navigation problem into a Brownian ratchet, which describes how drift of a particle can result from diffusion across an asymmetric landscape \cite{dill2010molecular, magnasco1993forced}. In the staPAW model, the asymmetry in movement direction mainly arise from the difference going up or down gradient around the T-state and average speed in the S-state.
}
  \label{fig:SI_BR}
\end{suppfigure}

\begin{suppfigure}[!t]
\centering
\includegraphics[width=.8\textwidth]{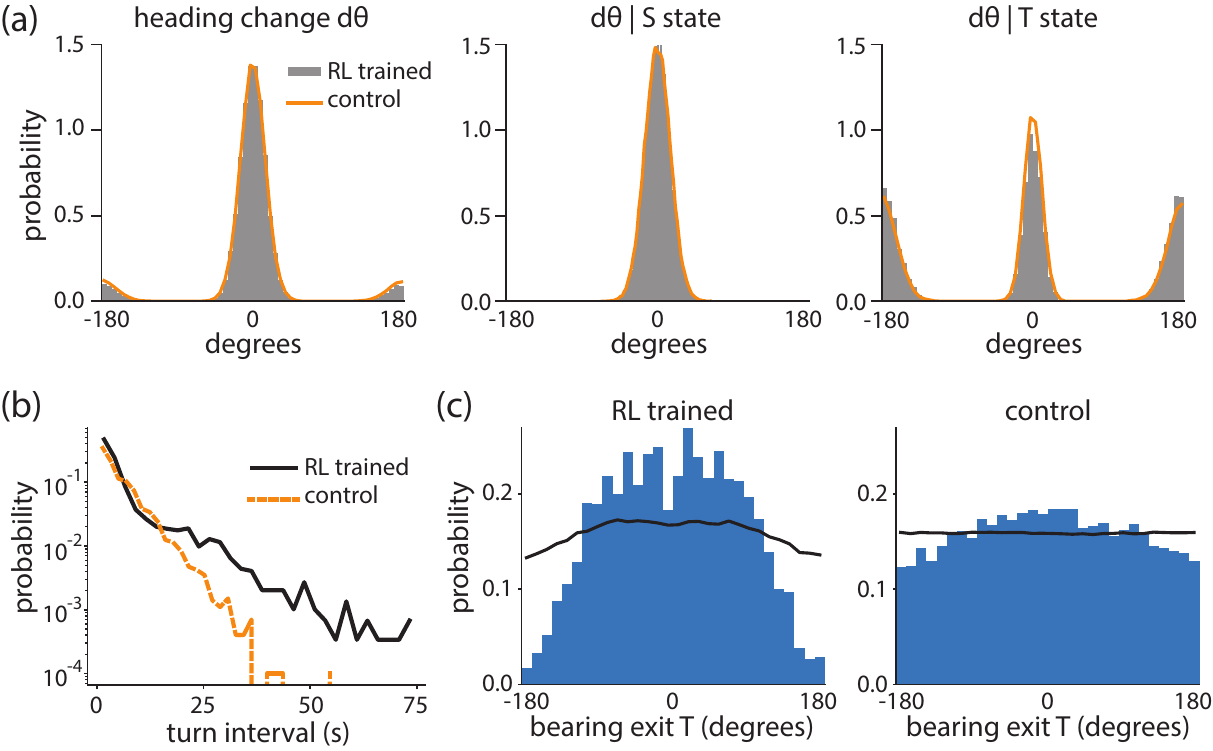}
\caption{Data-constrained RL model produces state-switching navigation strategies that qualitatively matches observations in worms. 
\textbf{(a)} The distribution of heading change $d\theta$ for all simulations (left), conditioned on S-state (middle), and conditioned on T-state (right), for both RL trained agents and a control model that does not have stimulus-dependent state transitions. 
\textbf{(b)} The turn interval distribution for RL trained model and control. \textbf{(c)} The distribution of bearing exiting T-state for RL trained (left) and control (right), with shuffled bearing differences shown in black line.
}
  \label{fig:SI_RL}
\end{suppfigure}


\end{document}